\documentclass[preprint,12pt]{elsarticle}
\usepackage{amsmath}
\usepackage{amsfonts}
\usepackage{amssymb}
\usepackage[utf8]{inputenc}
\usepackage[T2A]{fontenc}
\usepackage{bm}
\usepackage{float}
\usepackage{graphicx}

\begin{document}

\title{General solution of the Dirac equation with the Coulomb potential }

\author{A.A. Eremko$^1$, L.S. Brizhik$^1$, V.M. Loktev$^{1,2}$}
\address{$^1$ Bogolyubov Institute for Theoretical Physics of the National Academy of Sciences of Ukraine \\
Metrologichna Str., 14-b,  Kyiv, 03680, Ukraine \\
$^2$ National Technical University of Ukraine
“Igor Sikorsky Kyiv Polytechnic Institute” \\
Peremohy av., 37, Kyiv, 03056,  Ukraine}

\begin{abstract}
The Dirac equation with the Coulomb potential is studied. It is shown that there exists a new invariant in addition to the known Dirac and Johnson-Lippman ones. The solution of the Dirac equation, using the generalized invariant, and explicit expressions for the bispinors corresponding to the three sets of the invariants, their eigenvalues and quantum numbers are obtained. The general solution of the  Dirac equation with the Coulomb potential is shown to contain free parameters, whose variation transforms one particular solution into any other and controls spatial electron probability amplitude and spin polarization.  The electron probability densities and  spin polarizations are obtained in the general form and calculated explicitly for some electron states in the hydrogen-like energy spectrum. The spatial distributions of these characteristics  are shown to depend essentially on the invariant set, demonstrating physical difference of the states corresponding to different invariants.

\end{abstract}

\maketitle


\section{Introduction} 

Fundamental role of the Dirac equation (DE) became evident very soon after its discovery  \cite{Dirac1} (see \cite{Dirac,BetSol,LandauIV}). In particular, study of particle motion in the Coulomb potential in frame of the DE for hydrogen atom \cite{Darwin,Gordon} has shown the existence of the fine structure of its levels. Dirac equation was studied in numerous papers \cite{Davis,Martin,Biedenharn,Drake,Al-Hashimi}, whose basic results are summarised in textbooks on Quatum Mechaniscs (see, e.g., \cite{Dirac,BetSol,LandauIV,McConnell,Davydov,Messia,Bete}).

It has been shown that DE admits exact solutions \cite{Darwin} which means that Dirac problem with the Coulomb potential belongs to the class of integrable systems. It is known that quantum integrable systems can be characterized by the set of independent commuting operators $ \mathcal{H} = \lbrace \hat{H}_{1}, \hat{H}_{2}, \ldots \, , \hat{H}_{n} \rbrace $, where value $ n $ corresponds to the number of the degrees of freedom, and where the set $ \mathcal{H} $ includes the Hamiltonian and operators of the integrals of motion, i.e., invariants. Dirac theory has demonstrated that particle motion attains new features, so that such motion is characterized not only by spatial (orbital) coordinates, but also by new ones, spin variables. In this way notion of "eigen" (or intrinsic) momentum, spin, has been introduced.  It turned out that the set of particle vector states is represented by the tensor product of the corresponding spatial and spin spaces. Hence, integrability of the system means also existence of the full set of four mutually commuting operators.  
 
Worth to recall here that in the central-symmetric field integrals of motion are: the total angular momentum $ \mathbf{\hat{J}} = \hat{\mathbf{L}} \hat{I} + (\hbar /2) \bm{\hat{\Sigma}} $, which is given by the sum of the orbital $ \hat{\mathbf{L}} \hat{I} $ and spin $ \bm{\hat{\Sigma}} $ angular momenta, and introduced by Dirac operator $ \hat{{\cal I}}_{D} $ . Here $ \hat{I} $ is $ 4\times 4 $ unit matrix. This means that the set $ \mathcal{H}_{D} = \lbrace \hat{H}_{D}, \hat{J}^{2},\hat{J}_{z}, \hat{{\cal I}}_{D} \rbrace $ with $ \hat{H}_{D} $ been Dirac Hamiltonian, provides complete integrability of DE. This set allows to reduce initial system of four partial differential equations for amplitudes of the spinor field to the system of two first order ordinary differential equations for angular and radial functions of Dirac bispinor, similar to nonrelativistic quantum mechanics in which separation of variables reduces partial differential equations to ordinary ones. 

Like existence of symmetries, integrals of motion additional to the number of the degrees of freedom make the system degenerate which is manifested by degeneracy of the corresponding energy levels. An example of such degeneracy is particle motion in the field $ V \sim  1 /r $. In this case there are additional invariants, namely, the Laplace–Runge–Lenz vector in nonrelativistic theory (classical and quantum) and Johnson-Lippman operator \cite{John-Lip} for DE. Existence of the operator $ \hat{{\cal I}}_{JL} $ provides not only this degeneracy, called an \textit{accidental} one, but allows also to write down equations for radial functions in various equivalent representations (see above references). In \cite{Sukumar,JarvSted,Dahl} it has been shown that the methods of supersymmetric quantum mechanics can be used to obtain the complete energy spectrum and eigenfunctions of the DE. In this way the problem of hidden symmetries has been studied in details, while principally new meaning of the found solutions was left aside.  

In all these studies it has been shown that independent of the representation used, equations for radial functions are characterized by one hydrogen-like spectrum and one type of these functions in "normal modes" representation. Nevertheless, due to non-commutability of the operators  $ \hat{{\cal I}}_{JL} $ and $ \hat{{\cal I}}_{D} $,  there should exist eigenbispinors of DE corresponding to an alternative set $ \mathcal{H}_{JL} = \lbrace \hat{H}_{D}, \hat{J}^{2},\hat{J}_{z}, \hat{{\cal I}}_{JL} \rbrace $, different from those corresponding to the set $ \mathcal{H}_{D} $. Finding explicit form of such bispinors is the aim of the present paper. Also the differences in the spatial distributions of probability densities and average spin values corresponding to different sets of invariants, will be analysed.
   
It is very probable that particular quantum states of hydrogen atom corresponding to each set of invariants, can be realized depending on the chosen conditions of the given  experiment, under which the appearance or the value of the observable parameter is determined by matrix elements connected with external effects, such as electromagnetic fields, presence of other particles, etc. So one can hope that experimental technique will be developed, allowing to measure structure of atomic and molecular orbitals. In this respect we remind that the shape of Stark states of hydrogen atom have been already found \cite{PhysRev13}.

The paper is organized as follows. First, in Section 2, we will define explicit form of all invariants including the new one. Then, in Section 3, solution of DE for the so-called generalized invariant and expressions for the corresponding bispinors for each set of invariants including the given invariant, are presented. Here we also calculate  eigenvalues and determine their quantum numbers. Finally, in Section 4, the analysis is given of probability densities and spin polarizations for some of these states. The latter are shown to depend essentially on the invariant set they belong to. 

\section{Basic remarks} 

Wave DE  \cite{Dirac,BetSol,LandauIV}
\begin{equation}
\label{wDeq} 
i\hbar \frac{\partial \Psi}{\partial t} = \hat{H} \Psi , \quad \hat{H} = c \left( \hat{\mathbf{p}} - \frac{e}{c} \mathbf{A}(\mathbf{r}) \right) \bm{\hat{\alpha}} + e\varphi (\mathbf{r}) \hat{I} + m c^{2} \hat{\beta} 
\end{equation}
is the basis of the relativistic theory of electron in external electromagnetic field given by scalar $ \varphi (\mathbf{r}) $ and vector $ \mathbf{A}(\mathbf{r}) $ potentials. Here operator  $ \hat{H} $ is called Dirac Hamiltonian,   $ \Psi (\mathbf{r},t) = \left( \psi_{1}(\mathbf{r},t) \: \psi_{2}(\mathbf{r},t) \: \psi_{3}(\mathbf{r},t) \: \psi_{4}(\mathbf{r},t) \right)^{T} $ is four-component amplitude of spinor field (4-spinor, or bispinor),  $ c $ is speed of light ,  $ -e $ is electron charge,  $ \mathbf{\hat{p}} = -i\hbar \bm{\nabla} $ is momentum operator, $ \hat{\beta} $ and components $ \hat{\alpha}_{j} $ ($ j=x,y,z $) of the vector-matrix $ \bm{\hat{\alpha}} = \sum_{j} \mathbf{e}_{j} \hat{\alpha}_{j} $ together with the unit matrix  $ \hat{I} $ are $ 4\times 4 $ Hermitian Dirac matrices (DM). 
 
In time-independent fields, the states  $ \Psi (\mathbf{r},t) = \Psi (\mathbf{r}) \exp \left(-iEt/\hbar \right) $ with total energy $ E $ are determined by stationary DE 
\begin{equation}
\label{sDE}
\hat{H} \Psi = E \Psi .  
\end{equation}

It has been long ago shown by Darwin \cite{Darwin} that DE with the Coulomb potential created by point positive charge 
 $ Ze $, so that $ \mathbf{A} = 0,\: e\varphi (\mathbf{r}) = -Ze^{2}/r $ and 
\begin{equation}
\label{H_D}
\hat{H} = c \mathbf{\hat{p}} \bm{\hat{\alpha}} + V(r) \hat{I} + m c^{2} \hat{\beta} , \quad V(r) = - Ze^{2}/r ,
\end{equation}
admits exact solutions. The corresponding eigenfunctions describe states of both discrete and continuous spectra with the corresponding sets of quantum numbers. In their determination an important role belongs to the sets of physical parameters measurable simultaneously with the energy. In particular, for DE with the Hamiltonian (\ref{H_D}) such invariants are given by the components of the total angular momentum 
\begin{equation}
\label{J_tot}
\mathbf{\hat{J}} = \hat{\mathbf{L}} \hat{I} + \frac{\hbar}{2}  \bm{\hat{\Sigma}} \, .
\end{equation}
Here $ \hat{\mathbf{L}} = \mathbf{r}\times \mathbf{\hat{p}} $ is operator of the orbital momentum, and $ \left(\hbar /2 \right) \bm{\hat{\Sigma}} $ is operator of spin angular momentum, called "spin" for short. 

Since the components $ \mathbf{\hat{J}} $ do not commute with each other, usually operator of one of its component, namely $ \hat{J}_{z} $, and square of the operator of total angular momentum  $ \mathbf{\hat{J}}^{2} $ are choosen as the invariants. Dirac operator\footnote{Dirac operator usually is denoted as $ \hat{K} $, and Johnson-Lippman operator  (\ref{inv_J-L}) is denoted as $ \hat{A} $. Here we use Dirac matrices representation $ \hat{\rho}_{1} $, $ \hat{\rho}_{2} $, $ \hat{\rho}_{3} (\equiv \hat{\beta}) $ and $ \bm{\hat{\Sigma}} $, whose explicit form can be found  e.g. in  \cite{Dahl}} 
\begin{equation}
\label{inv_D} 
\hat{{\cal I}}_{D} = \hat{\rho}_{3} \left( \bm{\hat{\Sigma}} \cdot \hat{\mathbf{L}} + \hbar \right) 
\end{equation}
also commutes with the Hamiltonian (\ref{H_D})
 \cite{Dirac}. The full set of mutually commuting operators of $ \mathcal{H}_{D} $ determines eigenbispinors. 

In the Coulomb potential Johnson-Lippman operator  \cite{John-Lip} 
\begin{equation}
\label{inv_J-L} 
\hat{{\cal I}}_{JL} = \frac{mZe^{2}}{r} \bm{\hat{\Sigma}} \cdot \hat{\mathbf{r}} - \frac{i}{c} \hat{{\cal I}}_{D} \hat{\rho}_{1} \left( \hat{H} - mc^{2} \hat{\rho}_{3} \right) ,
\end{equation}
is also an integral of motion. It does not commute with  (\ref{inv_D}) which indicates the possibility to describe eigenstates of this Hamiltonian using eigenfunctions which correspond to the set  $ \mathcal{H}_{JL} = \lbrace \hat{H}, \hat{J}_{z}, \hat{J^{2}}, \hat{{\cal I}}_{JL} \rbrace $. 

It turns out that not only the operators (\ref{inv_D}) and (\ref{inv_J-L}) determine invariants commuting with the Hamiltonian $ \hat{H} $, but there exists another one, whose operator 
\begin{equation}
\label{A_BEL} 
\hat{{\cal I}}_{BEL}=\frac{1}{2i}\left[ \hat{{\cal I}}_{D}, \hat{{\cal I}}_{JL} \right] 
\end{equation}
commutes with $ \hat{H} $, but does not commute with either operator it is constructed of. Hence, the set 
$ \mathcal{H}_{BEL} = \lbrace \hat{H}, \hat{J}_{z}, \hat{J^{2}}, \hat{{\cal I}}_{BEL} \rbrace $ defines its own system of eigenfunctions, different from the previous one. 

It is easy to show that operators of the given invariants anticommute with each other 
\begin{equation}
\label{anticom} 
\lbrace \hat{{\cal I}}_{JL}, \hat{{\cal I}}_{D} \rbrace_{+} = \lbrace \hat{{\cal I}}_{BEL}, \hat{{\cal I}}_{D} \rbrace_{+} = \lbrace \hat{{\cal I}}_{BEL}, \hat{{\cal I}}_{JL} \rbrace_{+} = 0. 
\end{equation}
Each concrete set $ \lbrace \hat{H}, \hat{J^{2}}, \hat{J}_{z}, \hat{{\cal I}}_{inv} \rbrace $ corresponds to its state vectors $ \vert \varepsilon ,j, m_{j},\epsilon_{inv} \rangle $ with quantum numbers that are defined by equations 
\begin{equation}
\label{state} 
\begin{array}{c}
\hat{H}\vert \varepsilon ,j, m_{j},\epsilon_{inv} \rangle = mc^{2} \varepsilon \vert \varepsilon ,j,m_{j},\epsilon_{inv} \rangle , \\ 
\hat{J^{2}} \vert \varepsilon ,j,m_{j},\epsilon_{inv} \rangle = \hbar^{2} j\left(j + 1 \right) \vert \varepsilon ,j,m_{j},\epsilon_{inv} \rangle , \\
\hat{J}_{z} \vert \varepsilon ,j,m_{j},\epsilon_{inv} \rangle = \hbar m_{j} \vert \varepsilon ,j,m_{j},\epsilon_{inv} \rangle , \\
\hat{{\cal I}}_{inv} \vert \varepsilon ,j,m_{j},\epsilon_{inv} \rangle = \epsilon_{inv} \vert \varepsilon ,j,m_{j},\epsilon_{inv} \rangle .
\end{array}
\end{equation}
Here notation is used $ \varepsilon = E/mc^{2} $,  $ j $ is quantum number of total angular momentum, $ m_{j} $ is its projection on the polar axis,  $ \epsilon_{inv} $ is eigenvalue of the operator $ \hat{{\cal I}}_{inv} $.

While the quantum numbers that characterize discrete and continuous spectra, are determined by solutions of the DE, the eigenvalues of the operators $ \hat{J^{2}} $ and $ \hat{J}_{z} $ equal to $ \hbar^{2} j \left( j+1 \right) $ and $ \hbar m_{j} $, respectively, are defined algebraically from the commutation relations for the components of the operator  $ \mathbf{\hat{J}} $. Here  $ j $ takes positive half-integer values and  $ m_{j} $ are half-integer values in the interval $ -j \leq m_{j} \leq j $. Eigenvalues of the invariants can be obtained directly from the expressions for their squares  $ \hat{{\cal I}}_{D}^{2} $ \cite{Martin}, $ \hat{{\cal I}}_{JL}^{2} $ \cite{Dahl} and $ \hat{{\cal I}}_{BEL}^{2} $ that determine squares of the eigenvalues  $ \epsilon_{D}^{2} $, $ \epsilon_{JL}^{2} $ and $ \epsilon_{BEL}^{2} $. This gives the following expressions for the eigenvalues of the given invariants 
\begin{equation}
\label{eps_D} 
\epsilon_{D} = \pm \hbar \kappa_{j} ,\quad \kappa_{j} = j + 1/2 , 
\end{equation} 
\begin{equation}
\label{eps_JL} 
\epsilon_{JL} = \pm mZe^{2} a_{\varepsilon ,j} , \quad a_{\varepsilon ,j} = \sqrt{1 - \frac{\kappa_{j}^{2}}{Z^{2} \alpha^{2}} \left( 1 - \varepsilon^{2} \right)} ,
\end{equation}
\begin{equation}
\label{eps_BEL} 
\epsilon_{BEL} = \pm \hbar mZe^{2} \kappa_{j} a_{\varepsilon ,j} ,
\end{equation}
where $ \alpha = e^{2} /\hbar c $  is { \it{Sommerfeld fine structure constant}}.

Hence, one of the quantum numbers characterising vector states in (\ref{state}), is the number  
 $ \sigma_{inv} = \pm $ which determines signs of the eigenvalues of the corresponding invariant operator. Vector states $ \vert \varepsilon ,j,m_{j},\sigma_{inv} \rangle $ of different invariant sets correspond to different eigenbispinors of DE with different spatial properties. Since there exists more than one such set, one can introduce generalized invariant as an arbitrary linear combination 
\begin{equation}
\label{Inv_g} 
\hat{{\cal I}}_{gen} = \frac{c_{D}}{\hbar} \hat{{\cal I}}_{D} + \frac{c_{JL}}{mZe^{2}}  \hat{{\cal I}}_{JL} + \frac{c_{BEL}}{\hbar mZe^{2}} \hat{{\cal I}}_{BEL} ,
\end{equation}
where free constants  $ c_{D} $, $ c_{JL} $ and $ c_{BEL} $ determine weight contribution of the corresponding invariants. Similar to the case of quantum well potential studied in \cite{Annals16}, one can show that using generalized invariant in the set of mutually commuting operators inevitably leads to solutions with free parameters, so that their variation transforms one solution into any other. Explicit form of such generalized invariant and solution of DE with the Coulomb potential have not been studied before and are the main aims of the present paper. 

\section{Solution of Dirac Equation}

Equation (\ref{sDE}) is, in fact, a system of four equations for the corresponding components  $ \psi_{\nu} $ ($ \nu = 1,2,3,4 $), which acquire the explicit form after chosing DM. Below we use their conventional form and represent  $ 4 \times 4 $-dimensional matrices via $ 2 \times 2 $ ones.
In this case the Hamiltonian (\ref{sDE}) takes the form
\begin{equation}
\label{H_D-m} 
\hat{H}_{D} = \left( \begin{array}{cc}
\left( V(r)+mc^{2} \right) \hat{I}_{2} & c\bm{\hat{\sigma}} \mathbf{\hat{p}} \\ c\bm{\hat{\sigma}} \mathbf{\hat{p}} & \left( V(r)-mc^{2} \right) \hat{I}_{2}
\end{array}  \right) ,
\end{equation}
in which $ \hat{I}_{2} $ is a unit $ 2\times 2 $ matrix, and $ \hat{\sigma}_{j} $ ($ j=x,y,z $) are Pauli matrices,  $ \bm{\hat{\sigma}} = \sum_{j} \mathbf{e}_{j} \hat{\sigma}_{j} $. 

In this representation matrix operators of the invariants take the following forms: matrix of the total angular momentum reads as
\begin{equation}
\label{J_z} 
\hat{J}_{z} = \hat{L}_{z} \hat{I} + \frac{\hbar}{2} \hat{\Sigma}_{z} = \left( \begin{array}{cc}
\hat{L}_{z} \hat{I}_{2} + \frac{\hbar}{2} \hat{\sigma}_{z} & 0 \\ 0 & \hat{L}_{z} \hat{I}_{2} + \frac{\hbar}{2} \hat{\sigma}_{z}
\end{array}  \right) \, ,
\end{equation}
\begin{equation}
\label{J^2} 
\mathbf{\hat{J}}^{2} \equiv \hat{J^{2}} = \left( \begin{array}{cc}
\left( \mathbf{\hat{L}}^{2} + \frac{3}{4} \hbar^{2} \right) \hat{I}_{2} + \hbar \bm{\hat{\sigma}} \mathbf{\hat{L}} & 0 \\ 0 & \left( \mathbf{\hat{L}}^{2} + \frac{3}{4} \hbar^{2} \right) \hat{I}_{2} + \hbar \bm{\hat{\sigma}} \mathbf{\hat{L}}
\end{array}  \right) \, ,
\end{equation} 
Dirac invariant operator (\ref{inv_D}) reads as 
\begin{equation}
\label{el} 
\hat{{\cal I}}_{D} =  \left( \begin{array}{cc}
\hat{\Lambda} & 0 \\ 0 & - \hat{\Lambda} 
\end{array}  \right) \, , \quad \hat{\Lambda} = \bm{\hat{\sigma}} \hat{\mathbf{L}} + \hbar \hat{I}_{2} = \left( \begin{array}{cc}
\hat{L}_{z} + \hbar & \hat{L}_{x} - i\hat{L}_{y} \\ 
\hat{L}_{x} +i\hat{L}_{y} & - \hat{L}_{z} + \hbar   
\end{array} \right) ,
\end{equation} 
and Johnson-Lippman operator (\ref{inv_J-L}) reads as  
\begin{equation}
\label{MinvJ-L} 
\hat{{\cal I}}_{JL} = \left( \begin{array}{cc}
\bm{\hat{\sigma}}  \hat{\mathbf{A}}_{+} & i(Ze^{2}/cr) \hat{\Lambda} \\ -i(Ze^{2}/cr) \hat{\Lambda} & -\bm{\hat{\sigma}} \hat{\mathbf{A}}_{-} 
\end{array}  \right) \, .
\end{equation} 
Here 
\begin{equation}
\label{opPi} 
\hat{\mathbf{A}}_{\pm} = \frac{1}{2} \left( \hat{\mathbf{L}} \times \hat{\mathbf{p}} - \hat{\mathbf{p}} \times \hat{\mathbf{L}} \right) \pm mZe^{2} \frac{\mathbf{r}}{r} 
\end{equation}
is vector Laplace-Runge-Lenz operator. 

The explicit expression of the invariant (\ref{A_BEL}) can be obtained by direct calculation of the commutator which gives
\begin{equation}
\label{invB} 
\hat{{\cal I}}_{BEL} = \left( \begin{array}{cc}
 \bm{\hat{\sigma}} \hat{\mathbf{\cal{A}}}_{+} & \frac{Ze^{2}}{cr} \hat{\Lambda}^{2} \\ \frac{Ze^{2}}{cr} \hat{\Lambda}^{2} & \bm{\hat{\sigma}} \hat{\mathbf{\cal{A}}}_{-}
\end{array}  \right) , 
\end{equation} 
where $ \hat{\Lambda} $ has been defined above (see (\ref{el})), and the notatoion for the operator is used (cp. (\ref{opPi}))
\begin{equation}
\label{vecB} 
\hat{\mathbf{
\cal{A}}}_{\pm} = \frac{1}{2} \left( \hat{\mathbf{L}} \times \hat{\mathbf{A}}_{\pm} - \hat{\mathbf{A}}_{\pm} \times \hat{\mathbf{L}} \right) . 
\end{equation} 

Taking into account the block form of DM, one can represent the bispinor as
\begin{equation}
\label{bispinor} 
\Psi \left( \mathbf{r} \right) = \left( \begin{array}{c} \psi^{(u)} \left( \mathbf{r} \right)  \\ \psi^{(d)} \left( \mathbf{r} \right) 
\end{array} \right) , \quad 
\psi^{(u)} = \left( \begin{array}{c} 
\psi^{(1)} \left( \mathbf{r} \right)  \\ \psi^{(2)} \left( \mathbf{r} \right) 
\end{array} \right) , \; \psi^{(d)} = \left( \begin{array}{c}
\psi^{(3)} \left( \mathbf{r} \right)  \\ \psi^{(4)} \left( \mathbf{r} \right) 
\end{array} \right) .
\end{equation}
Here $ \psi^{(u/d)} $ are its upper/lower spinors, respectively, with the components  $ \psi^{(\nu )} $ and $ \psi^{(\nu + 1)} $, where  $ \nu =1 $ for the upper bispinor, and $ \nu = 3 $ for the lower one. 

In the central-cymmetric potential in all invariant sets, described above, operators $ \hat{H} $, $ \hat{J}_{z} $ and $ \hat{J^{2}} $ are common. Therefore, in all these cases bispinor  (\ref{bispinor}) ought to satisfy equations
\[
\hat{J^{2}} \Psi_{j,m_{j}} = \hbar^{2} j \left( j+1 \right) \Psi_{j,m_{j}} , \quad \hat{J}_{z} \Psi_{j,m_{j}} = \hbar m_{j} \Psi_{j,m_{j}}  ,
\]
whose solutions can be found in spherical coordinate system.

\subsection{Eigenbispinors of the invariant $ \hat{J}_{z} $} 

In spherical coordinate system one has $ \hat{L}_{z} = -i\hbar (\partial /\partial \varphi ) $. It follows from the matrix form (\ref{J_z}) that upper $ \psi^{(u)} $ and lower $ \psi^{(d)} $ spinors in the equality  $ \hat{J}_{z} \Psi_{m_{j}} = \hbar m_{j} \Psi_{m_{j}} $ satisfy the same equation: 
\begin{equation}
\label{EqJ_z} 
\hbar \left( \begin{array}{cc}
-i \frac{\partial}{\partial \varphi} + \frac{1}{2} & 0 \\ 0 & -i \frac{\partial}{\partial \varphi} - \frac{1}{2}  
\end{array}  \right) \left( \begin{array}{c} 
\psi^{(\nu)}_{m_{j}} \left( \mathbf{r} \right)  \\ \psi^{(\nu +1)}_{m_{j}} \left( \mathbf{r} \right) 
\end{array} \right) = \hbar m_{j} \left( \begin{array}{c} 
\psi^{{(\nu)}}_{m_{j}} \left( \mathbf{r} \right)  \\ \psi^{(\nu +1)}_{m_{j}} \left( \mathbf{r} \right) 
\end{array} \right) .
\end{equation} 
Its solutions are given by functions $ \psi^{(\nu )}_{m_{j}} (\mathbf{r}) = \exp \left(im_{\nu}\varphi \right)\psi^{(\nu )} \left(r,\vartheta \right) $, where $ \varphi ,\,\vartheta $ are azimuthal and polar angles, and integer numbers   $ m_{\nu} $ are connected via the relations 
\begin{equation}
\label{m_1,m_2} 
 m_{\nu} = m_{1} = m_{j} - 1/2 , \quad m_{\nu +1} = m_{2} = m_{j} + 1/2 .
\end{equation}
Thus, from equation  $ \hat{J}_{z} \Psi = \hbar m_{j} \Psi $ we obtain the dependence of the bispinor (\ref{bispinor}) components on angle $ \varphi $:
\begin{equation}
\label{eigenJ_z} 
\psi^{(u/d)} \left( \mathbf{r} \right) = \psi_{m_{j}}^{(u/d)} \left( \mathbf{r} \right) = \left( \begin{array}{c} 
e^{im_{1}\varphi}\psi^{(\nu)} \left( r,\vartheta \right) \\ e^{im_{2}\varphi}\psi^{(\nu +1)} \left( r,\vartheta \right) 
\end{array} \right) , \quad \begin{array}{c}
m_{1} = m_{j} - 1/2 , \\
m_{2} = m_{1} + 1 = m_{j} + 1/2 .
\end{array}
\end{equation}

\subsection{Eigenbispinors of the invariant $ \hat{J^{2}} $} 

The identity $ \left( \bm{\hat{\sigma}} \mathbf{\hat{L}} \right)\left( \bm{\hat{\sigma}} \mathbf{\hat{L}} \right) = \mathbf{\hat{L}}^{2} - \hbar \bm{\hat{\sigma}} \mathbf{\hat{L}} $  allows us to represent Legendre operator in the spinor form  \cite{Biedenharn} $ \mathbf{\hat{L}}^{2} = \hat{\Lambda}^{2} - \hbar \hat{\Lambda} $ and represent invariant  (\ref{J^2}) as 
\[
\hat{J^{2}} =  \left( \begin{array}{cc}
\hat{\Lambda}^{2} -  (\hbar^{2}/4) \hat{I}_{2}  & 0 \\ 0 & \hat{\Lambda}^{2} -  (\hbar^{2}/4) \hat{I}_{2}
\end{array}  \right) \, ,
\]
and equality $ \hat{J^{2}} \Psi_{j,m_{j}} = \hbar^{2} j \left( j+1 \right) \Psi_{j,m_{j}} $ leads to a single equation for upper and lower spinors, similar to the above, namely to the equation 
\begin{equation}
\label{Eq-J^2} 
\left( \hat{\Lambda}^{2} -( \hbar^{2}/4) \hat{I}_{2} \right) \psi^{(u/d)}_{j,m_{j}} \left( \mathbf{r} \right) = \hbar^{2} j \left( j+1 \right) \psi^{(u/d)}_{j,m_{j}} \left( \mathbf{r} \right)  
\end{equation} 
which includes the operator
\begin{equation}
\label{mLambda} 
\hat{\Lambda} = \left( \begin{array}{cc}
-i\hbar \frac{\partial}{\partial \varphi} + \hbar & \hbar e^{- i\varphi} \left( - \frac{\partial}{\partial \vartheta} + i \frac{\cos \vartheta }{\sin \vartheta} \frac{\partial}{\partial \varphi} \right) \\ 
\hbar e^{i\varphi} \left( \frac{\partial}{\partial \vartheta} + i \frac{\cos \vartheta }{\sin \vartheta} \frac{\partial}{\partial \varphi} \right) & i\hbar \frac{\partial}{\partial \varphi} + \hbar   
\end{array} \right) .
\end{equation} 

The operators of invariants (\ref{el}), (\ref{MinvJ-L}) and (\ref{invB}) also include operator $ \hat{\Lambda} $, therefore, it is possible to find the solutions of corresponding eigenvalue problems via eigenspinors of (\ref{mLambda}). This matrix includes variables $ \vartheta $ and $ \varphi $, only, and its eigenspinors, according to (\ref{EqJ_z}), ought to be of the form 
$$ \chi_{\lambda} \left( \vartheta ,\varphi \right) = \left( f^{(\nu)}(\vartheta)\exp \left(im_{1}\varphi \right), \: f^{(\nu +1)}(\vartheta) \exp \left(im_{2}\varphi \right) \right)^{T} . $$ 
The equation $ \hat{\Lambda} \chi_{\lambda} = \lambda \chi_{\lambda} $ (see Appendix) leads to two spherical harmonics spinors (\ref{sphsp+2})-(\ref{sphsp-2}), one of which corresponds to the positive eigenvalue $ \lambda_{+} = \hbar \left( l + 1 \right) >0 $, and the other one -- to the negative, $ \lambda_{-} = - \hbar l  <0 $. 

Spinors  (\ref{sphsp+2}) and (\ref{sphsp-2}) are orthonormalized and satisfy the following equations   \cite{LandauIV} 
\begin{equation}
\label{relat_chi} 
 \hat{\sigma}_{r} \chi_{ l,m_{j},+} = i \chi_{l+1,m_{j},-} , \quad \hat{\sigma}_{r} \chi_{ l,m_{j},-} = -i \chi_{l-1,m_{j},+} ,
\end{equation}
where
\begin{equation} 
\label{sigma_r} 
\hat{\sigma}_{r} = \bm{\hat{\sigma}} \mathbf{r} /r = \left( 
\begin{array}{cc}
\cos \vartheta & e^{-i\varphi} \sin \vartheta \\ e^{i\varphi} \sin \vartheta & -\cos \vartheta
\end{array} \right) , \quad \hat{\sigma}_{r}^{2} = \hat{I}_{2} 
\end{equation} 
is a unit matrix which anticommutes with  the operator  $ \hat{\Lambda} $. 

Equation (\ref{Eq-J^2}) allows two types of solution. When $ \psi^{(u/d)} \sim \chi_{l, m_{j},+ } $ is chosen, Eq. (\ref{Eq-J^2}) is satisfied by spinor (\ref{sphsp+2}) with $ l = j - 1/2 $, i.e. $ \psi_{j,m_{j}}^{(u/d)} \sim \chi_{j-1/2, m_{j},+ } $. In the case $ \psi^{(u/d)} \sim \chi_{l, m_{j},- } $ it is necessary to put $ l = j + 1/2 $, hence, the solution is $ \psi_{j,m_{j}}^{(u/d)} \sim \chi_{j+1/2, m_{j},- } $. It is easy to note that 
\begin{equation}
\label{eigenspEq} 
\hat{\Lambda} \chi_{j\mp 1/2, m_{j},\pm} = \pm \hbar \left( j + \frac{1}{2} \right) \chi_{j\mp 1/2, m_{j},\pm} = \pm \hbar \kappa_{j} \chi_{j\mp 1/2, m_{j},\pm} . 
\end{equation}
Therefore, $ \chi_{j\mp 1/2, m_{j},\pm} $ are eigen spinors of $ \hat{\Lambda} $ with eigenvalues equal by absolute value $ \kappa_{j} = j + 1/2 $ (that belongs to the series of natural numbers) with opposite signs. 

In the general case the solution of Eq. (\ref{Eq-J^2}) is given by the linear combination 
\begin{equation}
\label{anzatsp} 
\begin{array}{c}
\psi_{j, m_{j}}^{(u)} = F^{(+)} \left( r \right) \chi_{j-1/2, m_{j},+}\left( \vartheta ,\varphi \right) + F^{(-)} \left( r \right) \chi_{j+1/2, m_{j},-}\left( \vartheta ,\varphi \right) , \\ 
\psi_{j, m_{j}}^{(d)} = G^{(+)} \left( r \right) \chi_{j-1/2, m_{j},+}\left( \vartheta ,\varphi \right) + G^{(-)} \left( r \right) \chi_{j+1/2, m_{j},-}\left( \vartheta ,\varphi \right) , 
\end{array} 
\end{equation}
where the coefficients at spherical harmonics spinors can depend on $ r $, only, and their upper indeces $ (\pm) $ indicate at which spinor, corresponding to positive or negative eigen value  $ \hat{\Lambda} $, this coefficient stands in upper $  F^{(\pm)} $ or lower $ G^{(\pm)} $ spinor. 

\subsection{Eigenbispinors of the Hamiltonian $ \hat{H} $} 

In representation  (\ref{H_D-m}) and (\ref{bispinor}) DE  is given by the following system of equations 
\begin{equation}
\label{DsysEq_1} 
\begin{array}{c}
\left(E + Ze^{2}/r - mc^{2} \right) \psi^{(u)} - c\bm{\hat{\sigma}} \mathbf{\hat{p}} \psi^{(d)}  = 0  , \\
- c\bm{\hat{\sigma}} \mathbf{\hat{p}}\psi^{(u)} + \left(E + Ze^{2}/r + mc^{2} \right) \psi^{(d)}  = 0  .
\end{array}
\end{equation}
To write down the above equations in polar coordinates, one can use the identity\footnote{This transformation, based on Pauli matrices algebra, is equivalent to transformations of DE using DM algebra \cite{McConnell,Martin}.} 
$$ 
\left( \bm{\hat{\sigma}} \mathbf{\hat{p}} \right) \left( \bm{\hat{\sigma}} \mathbf{r} \right) /r = \left( \mathbf{\hat{p}} \mathbf{r} - i \bm{\hat{\sigma}} \mathbf{\hat{L}} \right) \frac{1}{r} = \hat{p}_{r} - \frac{i}{r} \hat{\Lambda} ,
$$ 
where $ \hat{p}_{r} = -i\hbar \left( \partial /\partial r + 1/r \right) = \left(-i\hbar /r \right) \left( \partial /\partial r\right) r $  is Hermitian operator of the radial momentum  (projection of operator $ \mathbf{\hat{p}} $ on the direction $ \mathbf{r} $), and transform nondiagonal part in  system (\ref{DsysEq_1}) to the form  
\begin{equation}
\label{equiv2} 
\bm{\hat{\sigma}} \mathbf{\hat{p}} = \left( \bm{\hat{\sigma}} \mathbf{\hat{p}} \right) \hat{\sigma}_{r}^{2} = \left( \hat{p}_{r} - \frac{i}{r} \hat{\Lambda} \right) \hat{\sigma}_{r} .
\end{equation}

Substituting now spinors (\ref{anzatsp}) in DE and taking into account Eq. (\ref{equiv2}) and equalities (\ref{relat_chi}) which determine action of the matrix $ \hat{\sigma}_{r} $ on spherical spinors and which are eigenspinors of the operator  $ \hat{\Lambda} $, we get from Eq. (\ref{DsysEq_1}) the following equalities: 
\[
\begin{array}{c}
\left[ c\left( i\hat{p}_{r} + \frac{\hbar \kappa_{j}}{r} \right) F^{(-)} + \left( E + \frac{Ze^{2}}{r} + mc^{2} \right) G^{(+)}  \right] \chi_{j-1/2, m_{j},+} + \\
+ \left[ c\left( -i\hat{p}_{r} + \frac{\hbar \kappa_{j}}{r} \right) F^{(+)} + \left( E + \frac{Ze^{2}}{r} + mc^{2} \right) G^{(-)}  \right] \chi_{j+1/2, m_{j},-} = 0 , \\
\left[ c\left( i\hat{p}_{r} + \frac{\hbar \kappa_{j}}{r} \right) G^{(-)} + \left( E + \frac{Ze^{2}}{r} - mc^{2} \right) F^{(+)}  \right] \chi_{j-1/2, m_{j},+} + \\ 
+ \left[ c\left( -i\hat{p}_{r} + \frac{\hbar \kappa_{j}}{r} \right) G^{(+)} + \left( E + \frac{Ze^{2}}{r} - mc^{2} \right) F^{(-)}  \right] \chi_{j+1/2, m_{j},-} = 0 .
\end{array}
\]
In view of the independence of spinors  $ \chi_{j-1/2, m_{j},+} $ and $ \chi_{j+1/2, m_{j},-} $, these relations are valid at zero values of their "coefficients", resulting to the differential equations for radial functions  $ F^{(\pm)}(r) $ and $ G^{(\pm )}(r) $. It is easy to see that this system of four equations reduces to two independent pairs of ordinary differential equations of the first order, one -- for functions  $ F^{(+)}(r) $ и $ G^{(- )}(r) $, and the second one -- for functions $ F^{(-)}(r) $ and $ G^{(+ )}(r) $:
\begin{equation}
\label{radEq} 
\begin{array}{c}
\mp \frac{\hbar c}{r} \frac{d}{dr}rF^{(\pm)} + \frac{\hbar c \kappa_{j}}{r} F^{(\pm)} + \left( E + \frac{Ze^{2}}{r} + mc^{2} \right) G^{(\mp)} = 0 , \\
\pm \frac{\hbar c}{r} \frac{d}{dr}rG^{(\mp)} + \frac{\hbar c \kappa_{j}}{r} G^{(\mp)} + \left( E + \frac{Ze^{2}}{r} - mc^{2} \right) F^{(\pm)} = 0 .
\end{array}
\end{equation}
Similar equations for radial functions are well known and their solutions can be found in many textbooks (e.g.,  \cite{LandauIV,McConnell,BetSol}). The corresponding solutions are exact and  expressed via hypergeometric functions \cite{Darwin,Gordon} or via generalized Laguerre polynomials   \cite{Davis,Dirac,BetSol,LandauIV,McConnell,Bete}).

Asymptotics of these functions depends on energy and at large distances $ r \rightarrow \infty $  is given by expressions
\begin{equation}
\label{asimpt} 
\begin{array}{c}
\mathit{R}^{(u/d)} \sim e^{-\varkappa r} , \quad \varkappa \sim \sqrt{m^{2}c^{4} - E^{2}} , \; \textrm{at} \; E^{2} < m^{2}c^{4} , \\
\mathit{R}^{(u/d)} \sim e^{\pm ikr} , \quad k \sim \sqrt{E^{2} - m^{2}c^{4}} , \; \textrm{at} \; E^{2} > m^{2}c^{4} ,
\end{array}
\end{equation}
where  $ \varkappa $ is spatial decrement, and  $ k $ is wavenumber. In the first case the above solution corresponds to discrete bound states, and in the second one it describes free states of continuous spectrum. 

Radial functions can be found  expanding solutions into power series (see, e.g., \cite{McConnell}). It is convenient to introduce functions   $ f^{(\pm)} = rF^{(\pm)} $ and $ g^{(\pm)} = rG^{(\pm)} $ and dimensionless coordinate  $ \xi = \left(mc   /\hbar \right) r $, where $ \hbar /mc $ is Compton wavelength of electron. This transforms Eqs.  (\ref{radEq}) into the system of equations for functions  $ f^{(\pm)}(\xi ) $ and $ g^{(\pm)}(\xi ) $.  The solutions in the discrete spectrum can be searched for in the form \cite{McConnell}
\begin{equation}
\label{f,g} 
\begin{array}{c}
f^{(\pm)}\left( \xi \right) = e^{-\varkappa \xi} u^{(\pm)} \left( \xi \right) , \quad 
u^{(\pm)} \left( \xi \right) =\xi^{\gamma } \sum_{n=0}^{\infty} b_{n}^{(\pm)} \xi^{n} , \\
g^{(\pm)} \left( \xi \right) = e^{-\varkappa \xi} v^{(\pm)} \left( \xi \right) ,\quad v^{(\pm)} \left( \xi \right) = \xi^{\gamma } \sum_{n=0}^{\infty} d_{n}^{(\pm)} \xi^{ n} ,
\end{array}
\end{equation}
where 
\begin{equation}
\label{k,a} 
\varkappa = \sqrt{ 1 - \varepsilon^{2} }  ,  \quad  \varepsilon = E/mc^{2}  < 1 
\end{equation} 
are dimensionless damping and energy, and value $\gamma $ accounts for the principal possibility of the existence of power series with non-integer powers.  

Substituting such functions into Eq.  (\ref{radEq}), we come to equations 
\begin{equation}
\label{Eq-u,v_sigm} 
\begin{array}{c}
\mp \frac{d u^{(\pm)}}{d \xi} + \left( \frac{ \kappa_{j} }{\xi} \pm \varkappa \right) u^{(\pm)} + \left( 1 + \varepsilon + \frac{Z\alpha}{\xi} \right) v^{(\mp)} = 0 , \\

\mp \frac{d v^{(\mp)}}{d \xi} - \left( \frac{ \kappa_{j} }{\xi} \mp \varkappa \right) v^{(\mp)} + \left( 1 - \varepsilon - \frac{Z\alpha}{\xi} \right) u^{(\pm)} = 0 ,
\end{array}
\end{equation} 
in which $ \alpha $ is the fine structure constant introduced in (\ref{eps_JL}). For functions  $ u(\xi) $ and $ v(\xi) $ in the form of series  (\ref{f,g}), Eqs. (\ref{Eq-u,v_sigm}) transform to power expressions. Then one can see, that Eqs. (\ref{Eq-u,v_sigm}) can be satisfied provided coefficients at all powers of variable  $ \xi $ are equal zero \cite{McConnell} 
\begin{equation}
\label{xi_1} 
\begin{array}{c}
\left( \gamma \mp \kappa_{j} \right) b_{0}^{(\pm)} \mp Z\alpha d_{0}^{(\mp)} = 0 , \\
\pm Z\alpha b_{0}^{(\pm)} + \left( \gamma \pm \kappa_{j} \right) d_{0}^{(\mp)} = 0 
\end{array}
\end{equation}
for  $n=0$, and 
\begin{equation}
\label{xi_n} 
\begin{array}{c}
\left( n + 1 + \gamma \mp \kappa_{j} \right) b_{n+1}^{(\pm)} \mp Z \alpha d_{n+1}^{(\mp)} = \varkappa b_{n}^{(\pm)} \pm \left( 1 + \varepsilon \right) d_{n}^{(\mp)}  , \\
\pm Z\alpha b_{n+1}^{(\pm)} + \left( n + 1 + \gamma \pm \kappa_{j} \right) d_{n+1}^{(\mp)} = \pm \left( 1 - \varepsilon \right) b_{n}^{(\pm)} + \varkappa d_{n}^{(\mp)} 
\end{array}
\end{equation}
for all $n > 0 $. 

Equations (\ref{xi_1}) represent the system of \textit{homogenous} linear equations for coefficients $ b_{0}^{(\pm)} $, $ d_{0}^{(\mp)} $ which admit nontrivial solution provided the equality takes place
\begin{equation}
\label{mu} 
\gamma \equiv \gamma_{j} = \sqrt{\kappa^{2}_{j}  - Z^{2}\alpha^{2}} ,
\end{equation}
where the square root has positive sign, only, as it follows from the condition of the convergence of  normalization integral. This sign of $ \gamma $ in Eqs. (\ref{xi_1}) implies the relation
\begin{equation}
\label{d_1} 
\sqrt{\kappa_{j}  \mp \gamma_{j} } b_{0}^{(\pm)} + \sqrt{ \kappa_{j} \pm \gamma_{j} } d_{0}^{(\mp)} = 0  .
\end{equation}
Similar relations for coefficients  with $n>0$ follow from Eqs. (\ref{xi_n}) and definition $\varkappa $ (see (\ref{k,a})): 
\begin{equation}
\label{b_n-d_n} 
\left( \frac{n + \gamma_{j} \mp \kappa_{j} }{\sqrt{1+\varepsilon}} - \frac{Z\alpha}{\sqrt{1-\varepsilon}} \right) b_{n}^{(\pm )} \mp  \left( \frac{n + \gamma_{j} \pm \kappa_{j} }{\sqrt{1-\varepsilon}} + \frac{Z\alpha}{\sqrt{1+\varepsilon}} \right) d_{n}^{(\mp )} = 0 .
\end{equation}

Equations (\ref{xi_n}) as the system of  \textit{inhomogenous} linear equations allows to express the coefficients  $ b_{n+1}^{(\pm)} $ and $ d_{n+1}^{(\mp)} $ via  $ b_{n}^{(\pm)} $ and $ d_{n}^{(\mp)} $, respectively, and using equalities  (\ref{d_1}), (\ref{b_n-d_n}), we can write down recurrent relations 
\begin{equation}
\label{rec_u} 
\begin{array}{c}
b_{1}^{(\pm)} = \frac{ \left(1 + \gamma_{j} \pm \kappa_{j} \right) \varkappa + Z\alpha \left( 1 - \varepsilon \right)}{ 1 + 2 \gamma_{j} } \left( 1 \mp \sqrt{\frac{\left( \kappa_{j} \mp \gamma_{j} \right) \left(1+\varepsilon \right) }{\left( \kappa_{j}  \pm \gamma_{j} \right) \left(1-\varepsilon \right) }} \right) b_{0}^{(\pm)} ,\\ 
b_{n+1}^{(\pm)} = 2 \frac{ \left[ \left( n +1 + \gamma_{j} \pm \kappa_{j}  \right) \varkappa + Z\alpha \left( 1 - \varepsilon \right)\right] \left[ \left( n + \gamma_{j} \right) \varkappa - Z\alpha \varepsilon \right] }{ \left( n + 1 \right) \left( n +1 + 2 \gamma_{j} \right) \left[ \left( n + \gamma_{j} \pm \kappa_{j} \right) \varkappa + Z\alpha \left(1 - \varepsilon \right) \right] } b_{n}^{(\pm)} ,\quad \: n \geq 1 
\end{array}
\end{equation}
for the coefficients of the series $ u^{(\pm)} \left( \xi \right) $. Similarly, we get recurrent relations 
\begin{equation}
\label{rec_v} 
\begin{array}{c}
d_{1}^{(\mp)} = \frac{ \left( 1 + \gamma_{j} \mp \kappa_{j} \right) \varkappa - Z\alpha \left( 1 + \varepsilon \right) }{ 1 + 2 \gamma_{j} } \left( 1 \mp   \sqrt{\frac{\left( \kappa_{j} \pm \gamma_{j} \right) \left(1-\varepsilon \right) }{\left( \kappa_{j}  \mp   \gamma_{j} \right) \left(1+\varepsilon \right) }} \right) d_{0}^{(\mp)} ,\\ 
d_{n+1}^{(\mp)} = 2 \frac{ \left[ \left(n + 1 + \gamma_{j} \mp \kappa_{j}  \right) \varkappa - Z\alpha \left( 1 + \varepsilon \right)\right] \left[ \left( n + \gamma_{j} \right) \varkappa - Z\alpha \varepsilon \right] }{ \left( n + 1 \right) \left( n + 1 + 2 \gamma_{j} \right) \left[ \left( n + \gamma_{j} \mp \kappa_{j}  \right)\varkappa - Z\alpha \left(1 + \varepsilon \right) \right] } d_{n}^{(\mp)} , \quad \: n \geq 1 
\end{array}
\end{equation}
for the coefficients of the series $ v^{(\mp)} \left( \xi \right) $. The finiteness condition of these solutions  at $ \xi \rightarrow \infty $ requires series cut-off at certain  $ n = n_{r} $ so that the coefficients vanish  $ b_{n}^{(\pm)}=d_{n}^{(\mp)}=0$  for all $n>n_r$. This means that functions  $ u^{(\pm)}(\xi) $ and $ v^{(\mp)}(\xi) $ are polynomials. From Eqs. (\ref{rec_u}) and (\ref{rec_v}) we get that this takes place at given $ n_{r} \geq 1 $ when
$
\left(n_{r}  + \gamma _j\right) \varkappa = Z\alpha \varepsilon ,
$
from where we get expressions for the energies of the DE bound states (\ref{sDE}) 
\begin{equation}
\label{E_rel2} 
\varepsilon \equiv \varepsilon_{n_{r},j} = \frac{n_{r} + \gamma_{j}}{{\mathcal N}_{n_{r},j}} = 
 \sqrt{1 - \frac{Z^{2}\alpha^{2}}{{\mathcal N}^2_{n_{r},j}}} \equiv \sqrt{1 - \varkappa ^2 _{n_{r},j} }.\end{equation}
Here $\varkappa_{n_r,j}=Z\alpha /{\mathcal{ N}_{n_{r},j}}$ is their damping, and the notation  \begin{equation}
\label{K_n,j}
{\mathcal N}_{n_{r},j} = \sqrt{\left(n_{r} + \gamma_{j} \right)^{2} + Z^{2}\alpha^{2} }=
\sqrt{n^2_{r}+\kappa ^2_j+2n_r\gamma_j}
\end{equation} 
is used. 
 
Series cut-off at certain term $ n_{r} = 0 $ means that the coefficients  $ b_{n}^{(\pm)} = d_{n}^{(\mp)}=0 $ for all $n \geq 1$ at non-zero $ b_{0}^{(\pm)} $ and $ d_{0}^{(\mp)}$. Nevertheless, direct calculation shows that from relations (\ref{rec_u}) and (\ref{rec_v}) one such possibility follows, when $ b_{0}^{(+)} $ and  $ d_{0}^{(-)}$ are non-zero. Energies of the corresponding states,  $ \varepsilon _{0,j}= \sqrt{1 - Z^{2}\alpha^ {2}/\kappa^{2}_{j} } $, naturally, coincide  with  the value (\ref{E_rel2}) at $ n_{r} = 0 $. Similar condition for coefficients $ b_{1}^{(-)} $ and $ d_{1}^{(+)}$ can be satisfied at zero values $ b_{0}^{(-)} = d_{0}^{(+)} = 0 $, only, which means the absence of solutions of system (\ref{xi_1}) or its trivial solution.  

Therefore, solutions of Eqs. (\ref{Eq-u,v_sigm}) are given by functions $ u^{(\pm)} = \xi^{\gamma_{j}} P_{n_{r},j}^{(\pm)} $ and $ v^{(\mp)} = \xi^{\gamma_{j}} Q_{n_{r},j}^{(\mp)} $, and eigenbispinors of the discrete spectrum are given by the following expressions
\begin{equation}
\label{eig-bsp_A} 
\Psi_{n_{r}, j, m_{j} } (\mathbf{r}) = \frac{1}{r} \left( \begin{array}{c} 
e^{-\varkappa_{n_{r},j}\xi} \xi^{\gamma_{j}} \left( P_{n_{r},j}^{(+)} \chi_{j-1/2, m_{j},+} + P_{n_{r},j}^{(-)} \chi_{j+1/2, m_{j},-} \right)  \\ 
e^{-\varkappa_{n_{r},j}\xi} \xi^{\gamma_{j}} \left( Q_{n_{r},j}^{(+)} \chi_{j-1/2, m_{j},+} +  Q_{n_{r},j}^{(-)} \chi_{j+1/2, m_{j},-} \right)   
\end{array} \right) , 
\end{equation} 
in which spherical spinors $ \chi_{l, m_{j},\pm}\left(\vartheta ,\varphi \right) $ are defined in  Eqs.  (\ref{sphsp+2}), (\ref{sphsp-2}), polynomials  $ P_{n_{r},j}^{(\pm)} $ and $ Q_{n_{r},j}^{(\pm)} $ of the $ n_{r} $-th degree are given by recurrent Eqs.  (\ref{rec_u}), (\ref{rec_v}). The latter for the eigenvalues  (\ref{E_rel2}) take the form 
\begin{equation}
\label{P^+,-} 
\begin{array}{c}
P_{n_{r},j}^{(\pm)} \left( \xi \right) = \sum_{n=0}^{n_{r}} b^{(\pm)}_{n} \xi^{n}, \\
b_{1}^{(\pm)}\left(n_{r},j \right) = \mp \frac{ \left( \mathcal{N}_{n_{r},j} + 1 - n_{r} \pm \kappa_{j} \right) \left( \mathcal{N}_{n_{r},j} + n_{r} \mp \kappa_{j} \right) }{\left( 1+2\gamma_{j} \right) \left( \kappa_{j} \pm \gamma_{j} \right) \mathcal{N}_{n_{r},j} }  Z\alpha b_{0}^{(\pm)} , \\ 
b_{n+1}^{(\pm)}\left(n_{r},j \right) = 2 \frac{ \left( n + \mathcal{N}_{n_{r},j} + 1 - n_{r} \pm \kappa_{j} \right) \left( n - n_{r} \right)}{\left( n + 1 \right) \left( n + 1 + 2 \gamma_{j} \right) \left( n + \mathcal{N}_{n_{r},j} - n_{r} \pm \kappa_{j} \right)\mathcal{N}_{n_{r},j} } Z\alpha b_{n}^{(\pm)}\left(n_{r},j \right) , \quad n \geq 1 ;
\end{array}
\end{equation}
\begin{equation}
\label{Q^-,+} 
\begin{array}{c}
Q_{n_{r},j}^{(\mp)}\left( \xi \right) = \sum_{n=0}^{n_{r}} d^{(\mp)}_{n} \xi^{n} \\
d_{1}^{(\mp)}\left( n_{r},j\right) = - \frac{\left( \mathcal{N}_{n_{r},j} + n_{r} - 1 \pm \kappa_{j}\right) \left( \mathcal{N}_{n_{r},j} + n_{r} \mp \kappa_{j} \right) }{ \left( 1 + 2 \gamma_{j} \right) \left( \mathcal{N}_{n_{r},j} + n_{r} + \gamma_{j} \right) \mathcal{N}_{n_{r},j}} Z\alpha d_{0}^{(\mp)} , \\ 
d_{n+1}^{(\mp)}\left( n_{r},j \right) = 2 \frac{ \left( n + 1 - \mathcal{N}_{n_{r},j} - n_{r} \mp \kappa_{j} \right) \left(n - n_{r} \right) }{ \left( n + 1 \right) \left( n + 1 + 2 \gamma_{j} \right) \left(n - \mathcal{N}_{n_{r},j} - n_{r} \mp \kappa_{j} \right) \mathcal{N}_{n_{r},j} } Z\alpha d_{n}^{(\mp)}\left( n_{r},j\right) ,\quad n \geq 1 .
\end{array}
\end{equation}
It follows from the above that they are composed of two independent pairs,  $ P_{n_{r},j}^{(+)} $, $ Q_{n_{r},j}^{(-)} $, and  $ P_{n_{r},j}^{(-)} $, $ Q_{n_{r},j}^{(+)} $, so that each pair represents the solution of its 'own' system of Eqs.  (\ref{Eq-u,v_sigm}). It is important that coefficients of the power series in each pair are connected by relations (\ref{d_1}), (\ref{b_n-d_n}). Moreover, all coefficients are expressed via the first ones, hence, only the relation  (\ref{d_1}) can be considered, which gives
\begin{equation}
\label{d_1-b_1} 
d_{0}^{(-)} = - \sqrt{\frac{\kappa_{j} - \gamma_{j}}{\kappa_{j} + \gamma_{j}}} b_{0}^{(+)}  , \quad d_{0}^{(+)} = - \sqrt{\frac{\kappa_{j} + \gamma_{j}}{\kappa_{j} - \gamma_{j}}} b_{0}^{(-)}.
\end{equation} 
From the normalization condition for bispinors one gets equation for coefficients  $ b_{0}^{(\pm)} $ 
\begin{equation}
\label{norm_r} 
I \equiv \frac{\hbar}{mc} \int_{0}^{\infty} e^{-2\varkappa_{n_{r},j}\xi} \xi^{2\gamma_{j}} \left( \mid P_{n_{r},j}^{(+)} \mid^{2} + \mid P_{n_{r},j}^{(-)} \mid^{2} + \mid Q_{n_{r},j}^{(+)} \mid^{2} + \mid Q_{n_{r},j}^{(-)} \mid^{2} \right) d\xi = 1 .
\end{equation} 
It is convenient to split integral (\ref{norm_r}) into two terms, $ I = I_{+} + I_{-} $, where  
\[
I_{\pm} = \frac{\hbar}{mc} \int_{0}^{\infty} e^{-2\varkappa_{n_{r},j}\xi} \xi^{2\gamma_{j}} \left( \mid P_{n_{r},j}^{(\pm)} \mid^{2} + \mid Q_{n_{r},j}^{(\mp)} \mid^{2} \right) d\xi .
\]

Polynomials $ P_{n_{r},j}^{(\pm)} $ and $ Q_{n_{r},j}^{(\mp)} $ satisfy Eqs. (\ref{Eq-u,v_sigm}) at eigenvalues (\ref{E_rel2}), and one can show that the corresponding solutions are expressed via generalized Laguerre polynomials:  
\begin{equation}
\label{P^+,Q^-_L} 
\begin{array}{c}
P_{n_{r},j}^{(+)} \left( r_{n_{r},j} \right) = c_{1} \left( \mathit{L}_{n_{r}}^{2\gamma_{j}} \left( r_{n_{r},j} \right)  - \frac{n_{r} + 2\gamma_{j}}{\mathcal{N}_{n_{r},j}+\kappa_{j}} \mathit{L}_{n_{r}-1}^{2\gamma_{j}} \left( r_{n_{r},j} \right) \right) , \\
Q_{n_{r},j}^{(-)}\left( r_{n_{r},j} \right) = -c_{1} \sqrt{\frac{1 - \varepsilon_{n_{r},j}}{1 + \varepsilon_{n_{r},j}}} \left( \mathit{L}_{n_{r}}^{2\gamma_{j}} \left( r_{n_{r},j} \right) + \frac{n_{r} + 2\gamma_{j}}{\mathcal{N}_{n_{r},j}+\kappa_{j}} \mathit{L}_{n_{r}-1}^{2\gamma_{j}} \left( r_{n_{r},j} \right) \right) ,
\end{array} 
\end{equation}
\begin{equation}
\label{P^-,Q^+_L} 
\begin{array}{c}
P_{n_{r},j}^{(-)} \left( r_{n_{r},j} \right) = c_{2} \left( \frac{n_{r} }{\mathcal{N}_{n,j}+\kappa_{j}} \mathit{L}_{n_{r}}^{2\gamma_{j}} \left( r_{n_{r},j} \right)  -  \mathit{L}_{n_{r}-1}^{2\gamma_{j}} \left( r_{n_{r},j} \right) \right) , \\ 
Q_{n_{r},j}^{(+)}\left( r_{n_{r},j} \right) = c_{2} \sqrt{\frac{1 - \varepsilon_{n_{r},j}}{1 + \varepsilon_{n_{r},j}}} \left( \frac{n_{r} }{\mathcal{N}_{n_r,j}+\kappa_{j}} \mathit{L}_{n_{r}}^{2\gamma_{j}} \left( r_{n_{r},j} \right) + \mathit{L}_{n_{r}-1}^{2\gamma_{j}} \left( r_{n_{r},j} \right) \right) ,
\end{array}  
\end{equation}
where $ c_{1} $ and $ c_{2} $ are independent constants, which play the role similar to that of coefficients  $ b^{(\pm)}_{0} $ in Eqs.(\ref{P^+,-}), (\ref{Q^-,+}), and the notation
\begin{equation}
\label{x_n,j} 
 r_{n_{r},j} \equiv 2\varkappa_{n_{r},j} \xi = \frac{2Zr}{r_{B}\mathcal{N}_{n_{r},j}} 
\end{equation} 
is used, in which $ r_{B} = \hbar^{2} /me^{2} $ is Bohr radius. Expressions (\ref{P^+,Q^-_L}), (\ref{P^-,Q^+_L}) can be considered as the transition to the normal-mode representation \cite{Gordon,Drake,Dahl}.

Now, using expressions (\ref{P^+,Q^-_L}) and (\ref{P^-,Q^+_L}), we calculate the integrals 
\[
\mathit{I}_{+} = \mid c_{1} \mid^{2} \frac{r_{B}}{Z\left( 2\varkappa_{n_{r},j} \right)^{2\gamma_{j}} } \frac{2\mathcal{N}_{n_r,j}^{2} \Gamma \left( n_r + 2\gamma_{j} + 1 \right) }{\left( 1 + \varepsilon_{n_{r},j} \right) \left( \mathcal{N}_{n_r,j} + \kappa_{j} \right)  n_{r}! } ,
\]
\[
\mathit{I}_{-} = \mid c_{2} \mid^{2} \frac{r_{B} }{Z\left( 2\varkappa_{n_{r},j} \right)^{2\gamma_{j}} } \frac{2\mathcal{N}_{n_r,j}^{2} \Gamma \left( n_r + 2\gamma_{j} \right) }{\left( 1 + \varepsilon_{n_{r},j} \right) \left( \mathcal{N}_{n_r,j} + \kappa_{j} \right)  (n_{r}-1)! } .
\]
Their sum in the normalization condition (\ref{norm_r}) leads to the condition which determines constants  $ c_{1} $ and $ c_{2} $. Introducing now the common normalization constant
\begin{equation}
\label{cnrj} 
\sqrt{\frac{2Z}{r_{B} \mathcal{N}_{n_r,j}}} C_{n_{r},j} , \quad C_{n_{r},j} =  \sqrt{\frac{\left( 1 + \varepsilon_{n_{r},j} \right) \left( \mathcal{N}_{n_r,j} + \kappa_{j} \right) n_{r}! }{4\mathcal{N}_{n_r,j} \Gamma \left( n_r + 1 + 2\gamma_{j} \right)}}, 
\end{equation}
we can parametrise the constants $ c_{1} $ and $ c_{2} $ as 
\[
c_{1} = \left( 2\varkappa_{n_{r},j} \right)^{\gamma_{j}} C_{n_{r},j} \beta_{1} , \quad c_{2} = \left( 2\varkappa_{n_{r},j} \right)^{\gamma_{j}} \sqrt{\frac{n_{r} + 2\gamma_{j}}{n_{r} }} C_{n_{r},j} \beta_{2}  
\]
and transform condition (\ref{norm_r}) to the  equation
\begin{equation}
\label{norm2} 
\mid \beta _1 \mid^2 + \mid \beta _2\mid^2 = 1 . 
\end{equation}
It is easy to see that this condition leaves the ambiguity in the choice of the parameters $ \beta_{1} $ and $ \beta_{2} $, so that every choice $ \beta _1^{(I)} = \beta_{1} ;\: \beta _2^{(I)} = \beta_{2} $ can be juxtapose with another one $ \beta _1^{(II)} = -\beta_{2}^{\ast} ;\: \beta _2^{(II)} = \beta_{1}^{\ast} $. Two mutually orthogonal bispinors correspond to this alternative, both belonging to the same system of eigenbispinors of DE. These two possibilities can be distinguished by signs, e.g., by introducing the number $ \sigma = \pm $ in such a way, that 
\begin{equation}
\label{par-c_s} 
\begin{array}{ccc}
\sigma = + & \Rightarrow & \beta_1^{(+)}( \theta , \phi ) = e^{i\phi} \cos \theta , \quad \beta_2^{(+)}( \theta , \phi )  = e^{-i\phi} \sin \theta ; \\
\sigma = - & \Rightarrow & \beta_1^{(-)}( \theta , \phi ) = - e^{i\phi} \sin \theta , \quad \beta_2^{(-)}( \theta , \phi ) = e^{-i\phi} \cos \theta ,
\end{array}
\end{equation}
where $ \theta $ and phase  $ \phi $ are free, spin, parameters, and normalization condition (\ref{norm2}) is satisfied automatically.

Defined in such a way, coefficients allow to assign index $ \sigma $ to two orthonormalized bispinors:
\begin{equation}
\label{Bispin_gen} 
\Psi_{n_{r}, j, m_{j},\sigma } (\mathbf{r}) = \left( \frac{2Z}{r_{B}\mathcal{N}_{n,j}} \right)^{3/2} C_{n_{r},j} e^{-\frac{r_{n_{r},j}}{2 }} r_{n_{r},j}^{\gamma_{j}-1} \left( \begin{array}{c} 
 \tilde{\psi}^{(u)}_{n_{r}, j, m_{j},\sigma } \left(r_{n_{r},j} ,\vartheta ,\varphi \right)   \\ 
\sqrt{\frac{1 - \varepsilon_{n_{r},j}}{1 + \varepsilon_{n_{r},j}}} \tilde{\psi}^{(d)}_{n_{r}, j, m_{j},\sigma } \left(r_{n_{r},j} ,\vartheta ,\varphi \right)    
\end{array} \right) .
\end{equation}
where spinors 
\begin{equation}
\label{psi^u,d} 
\begin{array}{c}
\tilde{\psi}^{(u)}_{n_{r}, j, m_{j},\sigma } = \beta_1^{(\sigma )} \tilde{P}_{n_{r},j}^{(+)}\left( r_{n_{r},j} \right) \chi_{j-1/2, m_{j},+}\left(\vartheta ,\varphi \right) + \beta_2^{(\sigma )}  \tilde{P}_{n_{r},j}^{(-)}\left( r_{n_{r},j} \right) \chi_{j+1/2, m_{j},-} \left(\vartheta ,\varphi \right) , \\
\tilde{\psi}^{(d)}_{n_{r}, j, m_{j},\sigma } = 
 \beta_2^{(\sigma )}  \tilde{Q}_{n_{r},j}^{(+)}\left( r_{n_{r},j} \right) \chi_{j-1/2, m_{j},+} \left(\vartheta ,\varphi \right) -  \beta_1^{(\sigma )}  \tilde{Q}_{n_{r},j}^{(-)}\left( r_{n_{r},j} \right) \chi_{j+1/2, m_{j},-}\left(\vartheta ,\varphi \right) 
\end{array} 
\end{equation}
with their common multiplier define upper and lower ("small") spinors of the bispinor (\ref{Bispin_gen}), respectively. Constant $ C_{n_{r},j} $ in expressions (\ref{Bispin_gen}) is defined in Eq. (\ref{cnrj}), coefficients  $ \beta_1^{(\sigma)} $ and $ \beta_2^{(\sigma)} $ -- in Eq. (\ref{par-c_s}), and dimesionless distance -- in (\ref{x_n,j}).  Thus, polynomials  (\ref{P^+,Q^-_L}) and (\ref{P^-,Q^+_L}) can be represented in the exact form
\begin{equation}
\label{tildP,Q} 
\begin{array}{c}
\tilde{P}_{n_{r},j}^{(+)} \left(  x  \right) = \mathit{L}_{n_{r}}^{2\gamma_{j}} \left( x \right)  - \frac{n_{r} + 2\gamma_{j}}{\mathcal{N}_{n_{r},j}+\kappa_{j}} \mathit{L}_{n_{r}-1}^{2\gamma_{j}} \left( x \right) , \\
\tilde{Q}_{n_{r},j}^{(-)}\left( x \right) = \mathit{L}_{n_{r}}^{2\gamma_{j}} \left( x \right) + \frac{n_{r} + 2\gamma_{j}}{\mathcal{N}_{n_{r},j}+\kappa_{j}} \mathit{L}_{n_{r}-1}^{2\gamma_{j}} \left( x \right)  , \\
\tilde{P}_{n_{r},j}^{(-)} \left( x \right) = \frac{\sqrt{n_{r}\left( n_{r} + 2\gamma_{j} \right) } }{\mathcal{N}_{n_r,j}+\kappa_{j}} \mathit{L}_{n_{r}}^{2\gamma_{j}} \left( x \right)  - \sqrt{\frac{n_{r} + 2\gamma_{j}}{n_{r}}} \mathit{L}_{n_{r}-1}^{2\gamma_{j}} \left( x \right)  , \\ 
\tilde{Q}_{n_{r},j}^{(+)}\left( x \right) = \frac{\sqrt{n_{r}\left( n_{r} + 2\gamma_{j} \right) } }{\mathcal{N}_{n_r,j}+\kappa_{j}} \mathit{L}_{n_{r}}^{2\gamma_{j}} \left( x \right) + \sqrt{\frac{n_{r} + 2\gamma_{j}}{n_{r}}} \mathit{L}_{n_{r}-1}^{2\gamma_{j}} \left( x \right)  ,
\end{array}
\end{equation}
where $ x \equiv r_{n_{r},j} $. 

Bispinors (\ref{Bispin_gen}) define general solution of DE and  describe the spectrum of bound states   (\ref{E_rel2}). They are eigenbispinors of the operator set   $ \lbrace \hat{H}, \hat{J^{2}}, \hat{J}_{z}, \hat{{\cal I}}_{gen} \rbrace $, where operator $ \hat{{\cal I}}_{gen} $ is defined in Eq. (\ref{Inv_g}). Vector states corresponding to (\ref{Bispin_gen}), are characterized by quantum numbers  $ n_{r} $, $ j $, $ m_{j} $ and $ \sigma $, where the last one is, in fact, $ \sigma_{gen} $, and defines the sign of the  $ \hat{{\cal I}}_{gen} $ eigenvalue. The square of the eigenvalue, as it can be easily checked, is expressed via eigenvalues of all invariants,  $ \epsilon_{gen}^{2} = c_{D}^{2} \kappa_{j}^{2} + c_{JL}^{2} a_{\varepsilon ,j}^{2} + c_{BEL}^{2} \kappa_{j}^{2} a_{\varepsilon ,j}^{2} $. 

At particular values of the free parameters, bispinors (\ref{Bispin_gen}) are eigenbispinors of one of the operator sets, which include invariants   (\ref{inv_D}), (\ref{inv_J-L}) or (\ref{A_BEL}). For instance, at $ \theta =\phi = 0 $  bispinors 
\begin{equation}
\label{bispin^D} 
\begin{array}{c} 
\Psi_{n_{r}, j, m_{j},+ }^{(D)} (\mathbf{r}) = \left( \frac{2Z}{r_{B}\mathcal{N}_{n,j}} \right)^{3/2} C_{n_{r},j} e^{-\frac{x_{n_{r},j}}{2 }} x_{n_{r},j}^{\gamma_{j}-1} 
\left( \begin{array}{c} 
 \tilde{P}_{n_{r},j}^{(+)} \chi_{j-1/2, m_{j},+}  \\ 
-\sqrt{\frac{1 - \varepsilon_{n_{r},j}}{1 + \varepsilon_{n_{r},j}}} \tilde{Q}_{n_{r},j}^{(-)} \chi_{j+1/2, m_{j},-}    
\end{array} \right) , \\
\Psi_{n_{r}, j, m_{j},- }^{(D)} (\mathbf{r}) = \left( \frac{2Z}{r_{B}\mathcal{N}_{n,j}} \right)^{3/2} C_{n_{r},j} e^{-\frac{x_{n_{r},j}}{2 }} x_{n_{r},j}^{\gamma_{j}-1}
\left( \begin{array}{c} 
\tilde{P}_{n_{r},j}^{(-)} \chi_{j+1/2, m_{j},-}  \\ 
\sqrt{\frac{1 - \varepsilon_{n_{r},j}}{1 + \varepsilon_{n_{r},j}}} \tilde{Q}_{n_{r},j}^{(+)} \chi_{j+1/2, m_{j},-}    
\end{array} \right) 
\end{array}
\end{equation}
are eigenbispinors of the set $ \lbrace \hat{H}_{D}, \hat{J}_{z}, \hat{J^{2}}, \hat{{\cal I}}_{D} \rbrace $ with $ \Psi_{n_{r}, j, m_{j},+ }^{(D)} $ corresponding to positive $ \hbar \kappa_{j} $ and $ \Psi_{n_{r}, j, m_{j},- }^{(D),} $  corresponding to negative $ - \hbar \kappa_{j} $ eigenvalues of Dirac invariant  (\ref{el}). In fact, they are Darwin solution (see \cite{Darwin,Gordon}).   Worth mentioning that the polynomials  $\tilde{P}_{n_{r},j}^{(\sigma)}$ and $\tilde{Q}_{n_{r},j}^{(-\sigma)}$ don't mix with one another in bispinors (\ref{bispin^D}), which makes their structure relatively simple as compared with the rest ones.   

Bispinor of the general solution (\ref{Bispin_gen}) can be represented in the form of the linear combination of Dirac solutions (\ref{bispin^D}), which play the role of DE "normal modes":
\begin{equation}
\label{eigbispi_g} 
\Psi_{n_{r}, j, m_{j},\sigma } (\mathbf{r})= \beta_1^{(\sigma)} \Psi_{n_{r}, j, m_{j},+ }^{(D)}(\mathbf{r}) + \beta_2^{(\sigma)} \Psi_{n_{r}, j, m_{j},- }^{(D)} (\mathbf{r}). 
\end{equation}

At the values $ \theta = \pi /4 $, $ \phi = -\pi /4 $ this bispinor takes the form 
\begin{equation}
\label{eigbispi_gJL} 
\Psi^{(JL)}_{n_{r}, j, m_{j},\sigma } (\mathbf{r})= \frac{1}{\sqrt{2}} \left( \sigma e^{-i\pi /4} \Psi_{n_{r}, j, m_{j},+ }^{(D)}(\mathbf{r}) + e^{i\pi /4} \Psi_{n_{r}, j, m_{j},- }^{(D)} (\mathbf{r})\right),  
\end{equation}
and one can see that it is the eigenbispinor of the Johnson-Lippman operator (\ref{MinvJ-L}) with eigenvalue  $ \sigma mZe^{2} a_{n_{r} ,j} $ (\ref{eps_JL}). Finally, at  $ \theta = \pi /4 $, $ \phi = 0 $ it is given by expression
\begin{equation}
\label{eigbispi_gBEL} 
\Psi^{(BEL)}_{n_{r}, j, m_{j},\sigma }(\mathbf{r}) = \frac{1}{\sqrt{2}} \left( \sigma \Psi_{n_{r}, j, m_{j},+ }^{(D)}(\mathbf{r}) + \Psi_{n_{r}, j, m_{j},- }^{(D)}(\mathbf{r}) \right) ,
\end{equation}
and, hence, is eigenbispinor of the operator  (\ref{invB}) with eigenvalue $ \sigma mZe^{2}\hbar \kappa_{j} a_{n_{r} ,j} $ (\ref{eps_BEL}). In other words, solution   (\ref{eigbispi_g}) in one case  corresponds to the set  $ \lbrace \hat{H}_{D}, \hat{J}_{z}, \hat{J^{2}}, \hat{{\cal I}}_{JL} \rbrace $, and in the second one -- to the set  $ \lbrace \hat{H}_{D}, \hat{J}_{z}, \hat{J^{2}}, \hat{{\cal I}}_{BEL} \rbrace $. Both operators,  $ \hat{{\cal I}}_{JL} $ and $ \hat{{\cal I}}_{BEL} $, anticommute with  $ \hat{{\cal I}}_{D} $ (see (\ref{anticom})), and expressions of their eigenbispinors via functions   $ \Psi_{n_{r}, j, m_{j},\pm }^{(D)} $ have the same form \cite{Dahl} except for the phase of the expansion coefficients (\ref{eigbispi_gJL}) and (\ref{eigbispi_gBEL}).

In the general case bispinor (\ref{Bispin_gen}) corresponds to the set of operators $ \lbrace \hat{H}_{D}, \hat{J}_{z}, \hat{J^{2}}, \hat{{\cal I}}_{gen} \rbrace $. The values of free parameters, angles   $ \theta $ and $ \phi $, depend on constants  $ c_{D} $, $ c_{JL} $ and $ c_{BEL} $ in Eq. (\ref{Inv_g}), varying which, as it has been mentioned above, any solution can be transformed into the other two, or, equivalently, any of the three solutions can be found from the general one.   

Eigen spectrum of the Dirac Hamiltonian is preserved and does not depend on the choice of the operator set $ \lbrace \hat{H}_{D}, \hat{J}_{z}, \hat{J^{2}}, \hat{{\cal I}}_{inv} \rbrace $, which determines the system of eigenbi\-spi\-nors. This is commonly known as {\it {random degeneration}}. According to the above, in addition to the known invariants, $ \hat{{\cal I}}_{D} $ and $ \hat{{\cal I}}_{JL} $, we have  generalized this phenomenon to the new invariant,  $ \hat{{\cal I}}_{BEL} $, extending it to $ \hat{{\cal I}}_{gen} $.

Among the states, characterized by quantum numbers  $ \lbrace n_{r},j,m_{j},\sigma \rbrace $, the states with  $ n_{r} = 0 $  are very special. It follows from the recurrent relations  (\ref{P^+,-}), (\ref{Q^-,+}) and can be directly seen from  (\ref{P^+,Q^-_L}), (\ref{P^-,Q^+_L}) that at $ n_{r} = 0 $ the only non-zero polynomilas are  $ \tilde{P}_{0,j}^{(+)} $ and $ \tilde{Q}_{0,j}^{(-)} $, while $ \tilde{P}_{0,j}^{(-)} = \tilde{Q}_{0,j}^{(+)} = 0 $. In these states eigenvalues of operators  $ \hat{{\cal I}}_{JL} $ and $ \hat{{\cal I}}_{BEL} $ also equal zero, sign  $ \sigma = \pm $ looses its meaning, and non-trivial solution for the Dirac invariant has positive sign, only. Hence, the states with $ n_{r} = 0 $ are described by eigenbispinors 
\begin{equation}
\label{Bisp_n=0} 
\Psi_{0, j, M,+ } (\mathbf{r}) = \left( \frac{2Z}{r_{B}\mathcal{N}_{0,j}} \right)^{3/2} C_{0,j} e^{-\frac{r_{0,j}}{2 }} r_{0,j}^{\gamma_{j}-1}
\left( \begin{array}{c} 
 \chi_{j-1/2, M,+}\left(\vartheta ,\varphi \right)  \\ 
-\sqrt{\frac{1 - \varepsilon_{0,j}}{1 + \varepsilon_{0,j}}} \chi_{j+1/2, M,-} \left(\vartheta ,\varphi \right)    
\end{array} \right) ,
\end{equation}
independent on the choice of eigenbispinors system. Expression of the states with $ n_{r} > 0 $ via bispinors (\ref{Bispin_gen}) with free parameter indicates spatial freedom in the choice of both charge and spin orientation distributions of particles occupying the corresponding levels. This question will be considered in the next Section.  

\section{Probability density and spin orientation of electron states } 

It is convenient to introduce the principal quantum number \cite{BetSol} $ n= n_{r} + \kappa_{j} = 1,2,... $, like it takes place in nonrelativistic theory of hydrogen atom, and to characterize the stationary states given by bispinors 
(\ref{Bispin_gen}) by the set of quantum numbers $ \lbrace n, j, m_{j}, \sigma \rbrace $ where the principal quantum number $ n $ takes positive integer values $ n = 1,2,\ldots $, quantum number of the total angular momentum $ j $ takes positive semi-integer values in the range $ 1/2 \leq j \leq n - 1/2 $ ($ 1 \leq \kappa_{j} \leq n $), magnetic quantum number $ m_{j} $ takes semi-integer values in the range $ -j \leq m_{j} \leq j $, and the number $ \sigma $ takes two values $ \sigma = \pm 1 $ except the states with the maximal value of $ j $ for the given $ n $, for $ j = n -1/2 $ ($ \kappa_{j} = n $) only one value  $ \sigma = + 1 $ is allowed. The radial quantum number $ n_{r} $ which determines the polynomials order in (\ref{Bispin_gen}), equals $ n_{r} = n - \kappa_{j} $. 

In this case the energies (\ref{E_rel2}) can be re-written in the form 
\begin{equation}
\label{E_N,j} 
\varepsilon_{n,j} = \sqrt{1 - \frac{Z^{2} \alpha^{2}}{\left( n-\Delta_{j} \right)^{2} + Z^{2}\alpha^{2} }} = \frac{n-\Delta_{j}}{\mathcal{N}_{n,j}} ,
\end{equation}
where the value
\begin{equation}
\label{delta_j} 
\Delta_{j} = \kappa_{j} - \sqrt{\kappa_{j}^{2} - Z^{2}\alpha^{2}} = \frac{Z^{2} \alpha^{2}}{\kappa_{j} + \sqrt{\kappa_{j}^{2} - Z^{2}\alpha^{2}}}, \quad \kappa_{j} = j + \frac{1}{2} 
\end{equation} 
characterizes \textit{fine} relativistic splitting of energy levels (\ref{E_N,j}) with given $ n $, and 
\begin{equation}
\label{calN_N,j} 
\mathcal{N}_{n,j} = \sqrt{\left( n-\Delta_{j} \right)^{2} + Z^{2}\alpha^{2}} = \sqrt{n^{2} - 2 \Delta_{j} \left( n - \kappa_{j} \right) } .
\end{equation} 

Eigenbispinors of the states with $ j = n - 1/2 $ ($ \kappa_{j} = n $), for which $ \mathcal{N}_{n,j} = n $ (see (\ref{calN_N,j})) and when the polynomials of the order $ n_{r} = n - \kappa_{j} = 0 $ are reduced to constants, are determined by the expression  (\ref{Bisp_n=0}). Their energy levels are given by expressions
\begin{equation}
\label{eps_N} 
\varepsilon_{n,n-1/2} \equiv \varepsilon_{n} = \sqrt{1 -Z^{2} \alpha^{2} / n^{2}} = \gamma_{n}/n , \quad  \gamma_{n} = \sqrt{n^{2} - \alpha^{2} Z^{2}}  
\end{equation}
and are  $ 2n $ times degenerate in accordance with the value of the magnetic quantum number  $ m_{j}=\pm 1/2, \ldots ,\ \pm (n-1)/2 $. The rest $ n-1 $  levels (at $ n \geq 2 $, $ j < (2n-1)/2 $) of the fine structure of each multiplet are described by spinors (\ref{Bispin_gen}) with polynomials of the order $ n_{r} = n - \kappa_{j} \geq 1 $, in which $ m_{j} $ takes $ 2(n-n_{r}) $ values, and $ \sigma $ takes both signs. Therefore, the corresponding level of the fine structure is $ 4(n-n_{r}) $ times degenerate, which means that $ 2n + 4\sum_{n_{r}=1}^{n-1} \left( n-n_{r} \right) = 2n^{2} $ states correspond to the same principal quantum number $ n $. 

Taking into account the explicit form of bispinors, it is easy to write down the square of the spinor field amplitude 
\begin{equation}
\label{rho_N} 
w_{n, j, m_{j},\sigma} = \Psi_{n, j, m_{j},\sigma }^{\dagger} (\mathbf{r}) \Psi_{n, j, m_{j},\sigma } (\mathbf{r}) = \mid \psi^{(u)}_{n, j, m_{j},\sigma} \mid^{2} + \frac{1 - \varepsilon_{n,j}}{1 + \varepsilon_{n,j}} \mid \psi^{(d)}_{n, j, m_{j},\sigma} \mid^{2} ,
\end{equation}
which characterises the distribution of the probability amplitude of electron presence on the given orbital, and to find charge spherical distribution  $ r^{2} \mid \Psi_{n, j, m_{j},\sigma } (\mathbf{r}) \mid^{2} $.

Another important and interesting characteristics of quantum states is electron spin orientation which is given by the meanvalue 
\begin{equation}
\label{vect-s} 
\Psi_{n, j, m_{j},\sigma }^{\dagger} (\mathbf{r}) \bm{\hat{\Sigma}} \Psi_{n, j, m_{j},\sigma } (\mathbf{r}) = \left. \psi^{(u)}\right.^{\dagger}_{n, j, m_{j},\sigma} \bm{\hat{\sigma}} \psi^{(u)}_{n, j, m_{j},\sigma} + \frac{1 - \varepsilon_{n,j}}{1 + \varepsilon_{n,j}} \left. \psi^{(d)}\right.^{\dagger}_{n, j, m_{j},\sigma} \bm{\hat{\sigma}} \psi^{(d)}_{n, j, m_{j},\sigma}   .
\end{equation}
It seems so far to our knowledge, the spatial dustribution of this value is unknown because it has not been studied before. 

In view of the smallness of the parameter $ Z\alpha \ll 1 $, contribution of the lower spinor of the order $ |Z\alpha |^2 \ll 1 $ in these expressions is negligibly small and can be ignored. By the same reason the upper spinor can be expanded with respect to the small parameter, and  terms $ \sim Z^{2}\alpha^{2} $ can be omitted. In fact, this corresponds to the nonrelativistic approximation when only upper spinor without account of relativistic corrections in it, is used to calculate the corresponding meanvalues. In this approximation the upper spinor is reduced to nonrelativistic Pauli one, 
\begin{equation}
\label{psi_u} 
\psi_{n,j,m_{j},\sigma}^{(P)} =  \beta_1^{(\sigma )} R_{n,j}^{(+)}\left( r \right) \chi_{j-1/2, m_{j},+}\left(\vartheta ,\varphi \right) + \beta_2^{(\sigma )}  R_{n,j}^{(-)}\left( r \right) \chi_{j+1/2, m_{j},-} \left(\vartheta ,\varphi \right)  ,
\end{equation}
where 
\[
R_{n,j}^{(\pm)} = \left(\frac{2Z}{\mathcal{N}_{n,j} r_{B}} \right)^{3/2} C_{n,j} e^{-r_{n,j}/2} r_{n,j}^{\gamma_{j} - 1} \tilde{P}_{n,j}^{(\pm)}(r_{n,j}),
\]
$r_{n,j}$ is the dimensionless radius (see (\ref{x_n,j})) and $ \mathcal{N}_{n,j} $ is determined in (\ref{calN_N,j}) (in view of the smallness of the parameter $ \alpha \ll 1 $ at not too large $ Z $ the value $ \mathcal{N}_{n,j} $ is numerically close to the principal quantum number value, $ \mathcal{N}_{n,j} \simeq n $). Finally, $ \tilde{P}_{n_{r},j}^{(\pm)} $ are polynomilas of the order $ n_{r} = n - \kappa_{j} $ and $ C_{n,j} $ is normalization constant determined in Eq. (\ref{cnrj}).

So we get that charge density is characterized by the function 
\begin{equation}
\label{w,s_nr} 
w_{n, j, m_{j},\sigma} \simeq \mid \psi^{(P)}_{n, j, m_{j},\sigma} \mid^{2} , 
\end{equation}
and spin orientation by the unit vector 
\begin{equation}
\label{spvec_nr} 
\mathbf{s}_{n, j, m_{j},\sigma} \simeq \frac{\left. \psi^{(P)}\right.^{\dagger}_{n, j, m_{j},\sigma} \bm{\hat{\sigma}} \psi^{(P)}_{n, j, m_{j},\sigma}}{w_{n, j,m_{j},\sigma}} .
\end{equation}

As it has been discussed above, the states on the levels with the value $ j $  maximal at the given principal quantum number  $ n $, i.e., at $ j = n - 1/2 $, are described by bispinors (\ref{Bisp_n=0}) in which $ \tilde{P}_{0,j}^{(-)} = \tilde{Q}_{0,j}^{(+)} = 0 $, which makes these states different from the rest ones. In particular, the ground level $ n=1 $ and states on the multiplets upper levels at $ n \geq 2 $ belong to such states. In the non-relativistic case they correspond to Pauli spinors
\[
\psi_{n,n-1/2,m_{j}}^{(P)} \left( \mathbf{r} \right)  =  \left(\frac{2Z}{n r_{B}} \right)^{3/2} \frac{1}{\sqrt{\left(2n \right)!} } e^{-r_{n}/2} r_{n}^{n - 1} \chi_{n-1, m_{j},+}\left(\vartheta ,\varphi \right)  .
\]
Probability density and spin orientation on the corresponding orbits are given by expressions
\[
w_{n, n-1/2, m_{j}} = \rho_{n}\left(r \right) \chi^{\dagger}_{n-1, m_{j},+} \chi_{n-1, m_{j},+} , \quad \rho_{n}\left(r \right) = \left(\frac{2Z}{n r_{B}} \right)^{3} \frac{1}{\left(2n \right)! } e^{-r_{n}} r_{n}^{2(n - 1)} , 
\]
\[
\mathbf{s}_{n, n-1/2, m_{j}} \left(\vartheta \right) = \chi^{\dagger}_{n-1, m_{j},+} \left(\vartheta ,\varphi \right) \bm{\hat{\sigma}} \chi_{n-1, m_{j},+}\left(\vartheta ,\varphi \right) ,
\]
from which it follows that both characteristics in these Pauli states are determined by spherical harmonics spinors  $ \chi_{n-1, m_{j},+} $ and radial functions. For example, two spinors correspond to the ground state   $ n=1 $,
\begin{equation}
\label{chi_0+} 
\chi_{0, \frac{1}{2},+} = \frac{1}{\sqrt{4\pi}} \left( \begin{array}{c} 1 \\
0
\end{array} \right) , \quad \chi_{0, -\frac{1}{2},+} = \frac{1}{\sqrt{4\pi}} \left( \begin{array}{c} 0 \\
1
\end{array} \right)  ,
\end{equation}
which gives $ w_{1, 1/2, \pm 1/2} = \rho_{1}\left(r \right)/4\pi $ and $ \mathbf{s}_{1, 1/2, \pm 1/2 } = \pm \mathbf{e}_{z} $.

The upper level  $ E_{2,3/2} $ of the doublet for $ n=2 $ is described by four spinors 
\[
\chi_{1, \frac{3}{2},+} = -i \sqrt{\frac{3}{8\pi}} e^{i\varphi} \sin \vartheta \left( \begin{array}{c} 1 \\
0
\end{array} \right) , \quad \chi_{1, -\frac{3}{2},+} = i \sqrt{\frac{3}{8\pi}} e^{-i\varphi} \sin \vartheta \left( \begin{array}{c} 0 \\
1
\end{array} \right) , 
\]
\[
\chi_{1, \frac{1}{2},+} = \frac{i}{\sqrt{8\pi}} \left( \begin{array}{c} 2 \cos \vartheta \\
-e^{i\varphi} \sin \vartheta
\end{array} \right) , \quad \chi_{1, -\frac{1}{2},+} = \frac{i}{\sqrt{8\pi}} \left( \begin{array}{c} e^{-i\varphi} \sin \vartheta \\
2 \cos \vartheta
\end{array} \right) , 
\]
from where it follows that
\[
w_{2,3/2,\pm 3/2} (\mathbf{r}) = \frac{3}{8\pi} \rho_{2}\left(r \right) \sin^{2} \vartheta ; \quad \mathbf{s}_{2,3/2,\pm 3/2} = \pm \mathbf{e}_{z},
\]
\[
w_{2,3/2,\pm 1/2} (\mathbf{r}) = \frac{1}{8\pi} \rho_{2}\left(r \right) \left( 1 + 3 \cos^{2} \vartheta \right) , \; \mathbf{s}_{2,3/2,\pm 1/2} = \pm \frac{\left( 5\cos^{2} \vartheta - 1 \right) \mathbf{e}_{z} - 2\sin 2\vartheta \mathbf{e}_{\rho} }{1 + 3\cos^{2} \vartheta} .
\]

Worth mentioning another peculiarity, discovered by Hartree  \cite{Hartree}, of the states on these levels, which complete population of the $ n $-th shell\footnote{The terminology is used when electrons with the principal quantum numbers  $ n = 1,2,3,\ldots $ are called electrons on the $ K-,L-,M-,\ldots $ shells \cite{LandauIII}.} of the hydrogen-like spectrum: summing probabilities  $ w_{n,j= n-1/2, m_{j}} $  by   $ m_{j} $ from $ m_{j} = +1/2 $ up to $ m_{j} = +j $ gives spherically symmetric charge distribution. In particular, one can see that $ w_{2,3/2,\pm 1/2} + w_{2,3/2,\pm 3/2} = \rho_{2}\left(r \right)/2\pi $.

The states on the rest sublevels of the multiplets with  $ n \geq 2 $ and $ j < n - 1/2 $ are characterized by additional quantity  $ \sigma = \pm $ and significantly depend on the invariant they correspond to. Darwin solution corresponds to the states with certain value of Dirac invariant and is described by bispinors  (\ref{bispin^D}). In Ref.  \cite{White} the probability amplitude (\ref{rho_N}) for this solution is given in  graphical presentation, and it is shown that in view of the fine structure constant smallness radial distributions, corresponding to Shr\"odinger and Dirac functions, are almost identical. This means that for Darwin solution the expression (\ref{w,s_nr}) with spinors (\ref{psi_u}) at $\beta_1^{(+)}=1,\, \beta_2^{(+)}=0,$ and $ \beta_1^{(-)}=0,\, \beta_2^{(-)}=1   $   reproduces Eq. (\ref{rho_N}) with high accuracy. Therefore, in this case the radial distribution of the electron cloud is also determined by the radial functions  $ R_{n,j}^{(\pm)} $, while angular dependence and spin orientation are described by spinors $ \chi_{j\mp1/2, m_{j},\pm} $. They are characterized by integer numbers $ l = j \pm 1/2 $, and although the number  $ l $ does not have physical meaning of the quantum number, characterizing eigenvalues of the operator $ L^{2} $ as an integral of motion, the states with $ l = 0,1,2,\ldots $ are conventionally marked by letters $ s,p,d,\ldots $. The states on the lower level  $ E_{2,1/2} $ of the doublet $ n=2 $ are given by Pauli spinors $ \psi_{2,1/2,m_{j},+}^{(Drw)} = R_{2,1/2}^{(+)} \chi_{0, m_{j},+} $ and $ \psi_{2,1/2,m_{j},-}^{(Drw)} = R_{2,1/2}^{(-)} \chi_{1, m_{j},-} $ and are defined as  $ 2s_{1/2} $ and $ 2p_{1/2} $ states, respectively. In them, neglecting corrections  $ \sim Z^{2}\alpha^{2} $, the radial functions 
\[
R^{(+)}_{2,1/2} = \frac{1}{\sqrt{2}} \left(\frac{Z}{r_{B}} \right)^{3/2} e^{-r_{2}/2} \left( 1 - \frac{1}{2} r_{2} \right) , \quad R^{(-)}_{2,1/2} = - \frac{1}{2\sqrt{6}} \left(\frac{Z}{r_{B}} \right)^{3/2} e^{-r_{2}/2} r_{2}   
\]
coincide with the radial functions of the Schr\"odinger equation, spinors $ \chi_{0, m_{j},+} $ are given in Eq. (\ref{chi_0+}), and 
\[
\chi_{1, \frac{1}{2},-} \left( \vartheta ,\varphi \right) = \frac{-i}{\sqrt{4\pi}} \left( \begin{array}{c} \cos \vartheta \\
e^{i\varphi} \sin \vartheta
\end{array} \right) , \quad \chi_{1, -\frac{1}{2},-} \left( \vartheta ,\varphi \right) = \frac{i}{\sqrt{4\pi}} \left( \begin{array}{c} -e^{-i\varphi} \sin \vartheta \\
\cos \vartheta
\end{array} \right) . 
\]
Taking these relations into account, we come to
\[
w^{(Drw)}_{2, 1/2, \pm 1/2,+} = \rho^{(Drw)}_{2,+}\left(r \right) = \left(\frac{Z}{r_{B}} \right)^{3} \frac{1}{8\pi} e^{-r_{2}} \left( 1 - \frac{1}{2} r_{2} \right)^{2} , \quad \mathbf{s}_{2, 1/2, \pm 1/2 ,+ } = \pm \mathbf{e}_{z} , 
\]
\[
w^{(Drw)}_{2, 1/2, \pm 1/2,-} = \rho^{(Drw)}_{2,-}\left(r \right) = \left(\frac{Z}{r_{B}} \right)^{3} \frac{1}{96\pi} e^{-r_{2}} r_{2}^{2} , \quad \mathbf{s}_{2, 1/2, \pm 1/2 ,- } = \pm \left( \cos \vartheta \mathbf{e}_{r} + \sin \vartheta \mathbf{e}_{\vartheta} \right) , 
\]
where $ \mathbf{e}_{r} $ and $ \mathbf{e}_{\vartheta} $ are orts in spherical coordinate system. From the given above probability densities we see, and this has been observed by White  \cite{White}, that in the states with certain value of Dirac invariant not only  $ s $-state, but also the state  $ 2p_{1/2} $ reveals spherical symmetry. 

Nonetheless, since the invariant itself is not determined \textit{a priori}, the states must be described by general solution, i.e., by bispinors   (\ref{Bispin_gen}) or (\ref{eigbispi_g}), or in the non-relativistic limit by Pauli spinors (\ref{psi_u})) with two free parameters  $ \left( \beta_1^{(\sigma )} , \beta_2^{(\sigma )} \right) $ that characterize spin degree of freedom which is manifested on these multiplet  sublevels. Thus, in the general case the states on the   $ E_{2,1/2} $ level are given by Pauli spinors 
\begin{equation}
\label{psi_2,1/2} 
\psi_{2,1/2,m_{j},\sigma}^{(P)} =  \beta_1^{(\sigma )} R_{2,1/2}^{(+)}\left( r \right) \chi_{0, m_{j},+} + \beta_2^{(\sigma )} R_{2,1/2}^{(-)}\left( r \right) \chi_{1, m_{j},-} \left(\vartheta ,\varphi \right) .
\end{equation}
and spherical symmetry of the probability density in these states is broken. Recall, if the sign  $ \sigma = + $ corresponds to pair  $ \left(\beta_{1},\,\beta_{2} \right) $, then the opposite sign  $ \sigma = - $ corresponds to -- $ \left(-\beta_{2}^{\ast},\,\beta_{1}^{\ast} \right) $ at $ \mid \beta_{1} \mid^{2} + \mid \beta_{2} \mid^{2} = 1 $. In this case not only spherical symmetry, but also symmetry with respect to reflection from the $ xy $-plane perpendicular to the polar axis, are broken.  This is due to the fact that Johnson-Lippman invariant operator (\ref{inv_J-L}), unlike Dirac operator (\ref{inv_D}), does not commute with the inversion operator, and, therefore, the general solution doesn't possess certain pairity.  The same conclusion is valid for the invariant $\hat{{\cal I}}_{BEL}$. 

Variation of the parameters  $ \beta_{1} $ and $ \beta_{2} $ in expression (\ref{psi_u}) determines  the deformation of electron cloud and changes of spin orientation which are mutually connected. In all cases the states with positive and negative values of the magnetic number $ m_{j} $ have the same probability density distribution, but opposite spin orientation. Taking into account the symbols used in quantum chemistry, expression (\ref{psi_2,1/2}) can be considered as $ sp $-hybrydization of the states on this level.

\begin{figure}[h!]
\begin{picture}(16,8)
  \includegraphics[width=4cm]{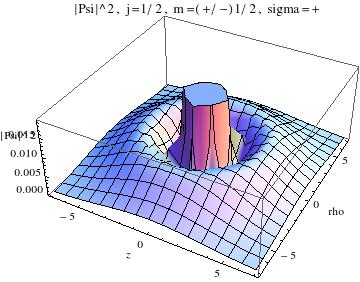}
   \includegraphics[width=4cm]{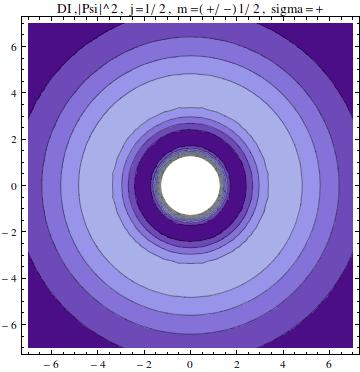}
    \includegraphics[width=1cm]{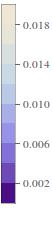}
  \includegraphics[width=4cm]{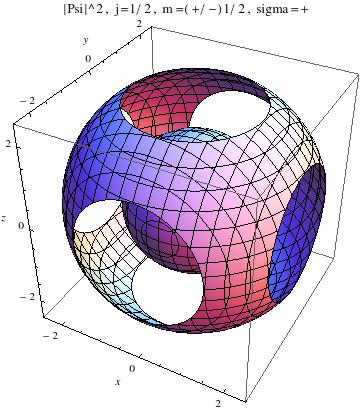}
  \end{picture}
  \caption{ Probability distribution for Dirac invariant, $n=2,\, j=1/2,\,m=\pm1/2,/, \sigma =+$: (a) probability distribution in the plane $(z,\rho)$; (b) isoenergy lines; (c) legend of Fig.1(b); (d)  probability distribution in the Cartesian coordinate system.}
  \label{fig:1}
\end{figure}

\begin{figure}[b]
\begin{picture}(16,8)
  \includegraphics[width=4cm]{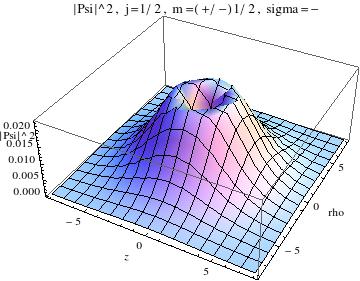}
   \includegraphics[width=4cm]{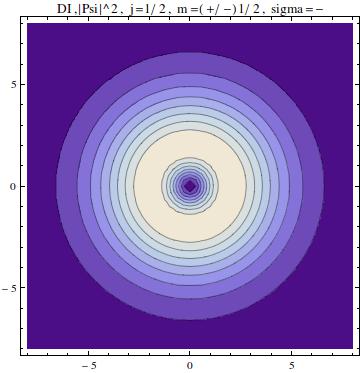}
    \includegraphics[width=1cm]{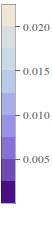}
  \includegraphics[width=4cm]{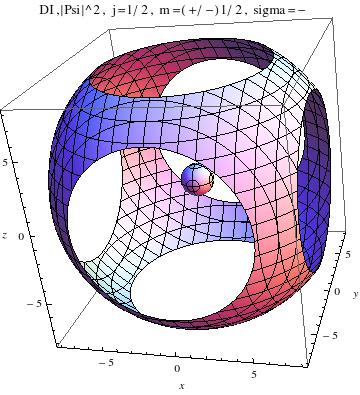}
  \end{picture}
  \caption{Probability distribution for Dirac invariant, $n=2,\, j=1/2,\, m=\pm1/2,\, \sigma=-$: (a) probability distribution in the plane $(z,\rho)$; (b) isoenergy lines; (c) legend of Fig.2(b); (d) probability distribution in the Cartesian coordinate system.}
  \label{fig:2}
\end{figure}

\begin{figure}[t]
\begin{picture}(16,8)
  \includegraphics[width=4cm]{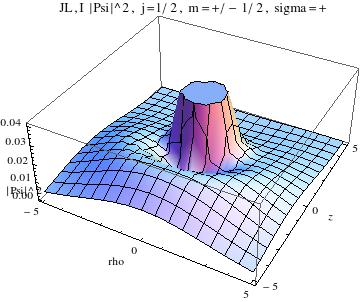}
   \includegraphics[width=4cm]{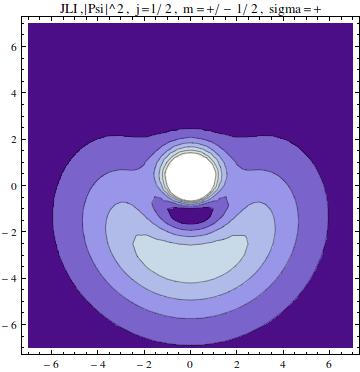}
    \includegraphics[width=1cm]{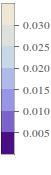}
  \includegraphics[width=4cm]{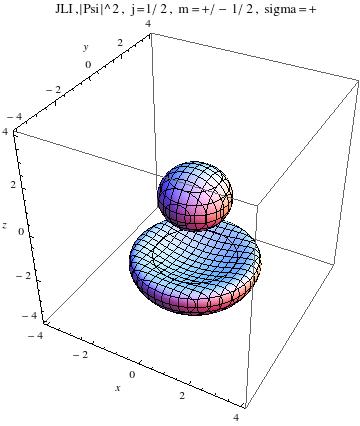}
  \end{picture}
  \caption{Probability distribution for Johnson-Lippman invariant, $n=2,\, j=1/2,\, m=\pm 1/2,\, \sigma =+$: (a) probability distribution in the plane $(z,\rho)$; (b)isoenergy lines; (c) legend to Fig.3(b); (d)  probability distribution in the Cartesian coordinate system.}
  \label{fig:3}
\end{figure}

\begin{figure}[t]
\begin{picture}(16,8)
  \includegraphics[width=4cm]{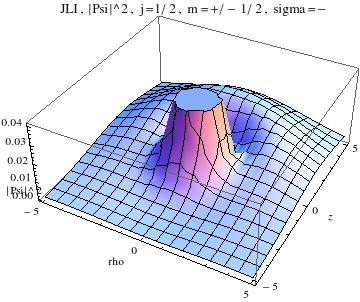}
   \includegraphics[width=4cm]{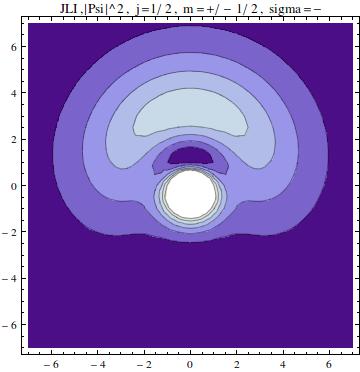}
    \includegraphics[width=1cm]{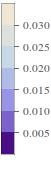}
  \includegraphics[width=4cm]{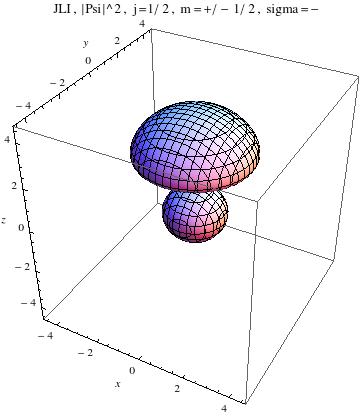}
  \end{picture}
  \caption{Probability distribution for Johnson-Lippman invariant, $n=2,\, j=1/2,\, m=\pm 1/2,\, \sigma =-$: (a) probability distribution in the plane $(z,\rho)$; (b) isoenergy lines; (c) legend to Fig.4(b); (d) probability distribution in the Cartesian coordinate system.}
    \label{fig:4}
\end{figure}

\begin{figure}[b]
\vspace{20mm}
\begin{picture}(16,8)
  \includegraphics[width=4cm]{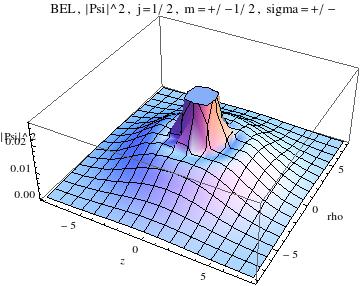}
   \includegraphics[width=4cm]{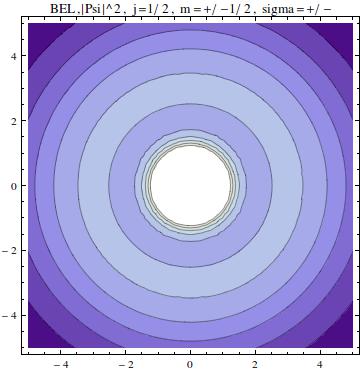}
    \includegraphics[width=1cm]{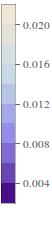}
  \includegraphics[width=4cm]{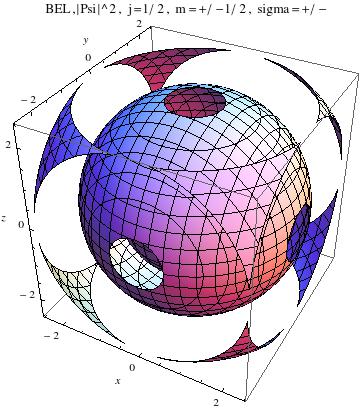}
  \end{picture}
  \caption{ Probability distribution for BEL invariant, $n=2,\, j= 1/2,\, m=\pm1/2,\, \sigma =\pm$: (a) probability distribution in the plane $(z,\rho)$; (b) isoenergy lines; (c) legend to Fig.1(b); (d)  probability distribution in the Cartesian coordinate system.}
  \label{fig:5}
\end{figure}

\begin{figure}[t]
\vspace{-15mm}
     \centering 
%\begin{picture}(5,5)
  \includegraphics[width=6cm]{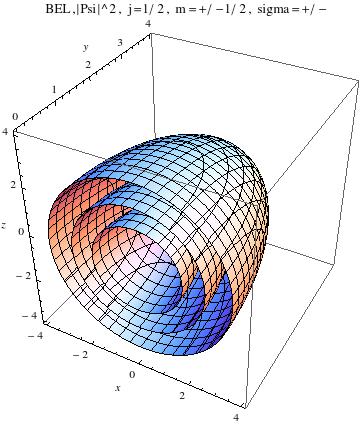}
%  \end{picture}
  \caption{Contour plot of the electron probability distribution in the Cartesian coordinate system for the probability value $ |\Psi_{2,1/2,\pm 1/2}^{(\pm)}|^2= 0.012$ for BEL invariant, $n=2,\, j=1/2,\, m=\pm 1/2,\, \sigma =\pm$.}
  \label{fig:6}
\end{figure}

Figures \ref{fig:1}--\ref{fig:6} show different presentations, described in figure captions, of the electron probability distributions for some energy states that correspond to different invariants, at the principal quantum number $n=2$. These figures demonstrate how the probability distribution changes depending on the spin invariant. On the other hand, spin distributions also significantly change and for each invariant are given by different expressions. 

\vspace{5cm}
%\vspace{1cm}

\newpage
\topmargin=-1cm
%\clearpage

\vspace{-2cm}

%\clearpage
\vspace{20mm}

%\clearpage

\clearpage

\section{Conclusions}
As it has been shown above, in the Coulomb potential there exist additional to the known before operator invariants for the DE  (namely, the Hamiltonian, Dirac or Johnson-Lippman invariant, square of the total angular momentum and one of its projections)  invariant $\hat{{\cal I}}_{BEL}$, determined in (\ref{A_BEL}), which commutes with the Hamiltonian, but does not commute with the operators it is constructed from.  The vector states corresponding to different invariant sets, are characterized by  different eigenbispinors of DE with different spatial properties and different spin polarizations. Since there exists more than one such set, there exists the generalized invariant in the form of a  linear combination (\ref{Inv_g}), and, therefore, the general solution of DE with the Coulomb potential contains free parameters, as we have  shown above (see (\ref{Bispin_gen}).  Variation of these free parameters  transforms one solution into any other. Moreover, the free parameters in the general solution of DE control not only spatial probability density, but also spin polarization of electrons, providing in such a way physical difference of the states corresponding to different invariants. 

Based on this study, one can formulate the following general statement: if a spin invariant has a diagonal block matrix structure, then the form of the spin-orbit interaction operator that is used in relativistic quantum mechanics, is preserved and is proportional to the product $  \bm{l} \cdot \bm{s}$. If the matrix of a spin invariant is not diagonal, as it takes place for spin invariants  $\hat{{\cal I}}_{JL}$ and  $\hat{{\cal I}}_{BEL}$, then the form of the spin-orbit interaction operator is different. Namely such situation takes place in the case of two-dimensional electron gas, for which there exist the cases when the spin-orbit interaction operator has the form different from the standard one, because the corresponding spin invariant matrix is non-diagonal.      

This together is a direct  manifestation of the spin states variability of the corresponding levels in the hydrogen-like spectrum. This can open new perspectives for spintronics devices, used in spincaloritronics, heat-to-spin conversion, energy harvesting, etc., for spin-orbit torque devices used in  electric magnetization switching, for design of quantum materials \cite{Naaman,Michaeli1,Chernyshov}, as well as for spin chemistry and relativistic chemistry \cite{Rosenberg,Bustami,Liu,Hore,Matysik}. A significant advantage of the studied in the present paper effect is the fact that it gives a possibility to govern spin states at minimal energy expenses. According to the obtained above results, energy of the different spin states at the same values of other quantum numbers are equal, and a small external factor, e.g., the one broking inversion symmetry, can be enough to cause electron transition from one spin orientation to another one. For instance, spin-orbit torque devices are not only much faster and more robust, but they also consume less power as compared to spin-transfer torque devices (see \cite{Li}). Another field  where this property is of potential importance, are biological systems, in which atoms with hydrogen-like spectrum are abandon. In these systems electrons play essential role not only in the  storage and transport of energy and information, but also in biorecognition connected with the spin selectivity and  chiral-induced spin selectivity (CISS), when electron transfer through diamagnetic material (macromolecules in the case of biological systems) results in  specific spin states of electrons  \cite{Michaeli,Varade,Zollner} which is attributed to spin-orbit interaction.

\vskip5mm 
{\bf Acknowledgement.} 
\textit{The work was carried out in the framework
of the budget program KPKVK~6541230 and the scientific program
0117U00236 of the Department of Physics and Astronomy of the
National Academy of Sciences of Ukraine.}

\appendix 

\section{Eigenspinors $ \hat{\Lambda} $} 

According to Eq. (\ref{eigenJ_z}), dependence of the bispinor components on the variable   $ \varphi $ is defined, and, therefore, solution of the spinor equation $ \hat{\Lambda} \chi_{\lambda} = \hbar \lambda \chi_{\lambda} $ can be searched for in the form
\begin{equation}
\label{chi_lambda} 
\chi_{\lambda}\left( \vartheta ,\varphi \right) = \left( \begin{array}{c} 
e^{im_{1}\varphi}f_{1} \left( \theta \right) \\ e^{im_{2}\varphi}f_{2} \left( \theta \right) 
\end{array} \right) , 
\end{equation}
where $ m_{1} = m_{j} - 1/2 $, $ m_{2} = m_{j} + 1/2 $, so that $ m_{2} - m_{1} = 1 $. The equation itself represents the system of equations for components  $ f_{\nu} \left( \theta \right) $, namely 
\begin{equation}
\label{systL} 
\begin{array}{c}
- \frac{d f_{2}\left( \vartheta \right)}{d \vartheta} - m_{2} \cot \vartheta f_{2}\left( \vartheta \right) = \left( \lambda - m_{2} \right) f_{1}\left( \vartheta \right) , \\
\frac{d f_{1}\left( \vartheta \right)}{d \vartheta} - m_{1} \cot \vartheta f_{1}\left( \vartheta \right) = \left( \lambda + m_{1} \right) f_{2}\left( \vartheta \right) .
\end{array}
\end{equation}
It is easy to see that this system  of the first order equations for the the two functions can be reduced to one equation of the second order for each function: 
\begin{equation}
\label{Eq-f(theta} 
\frac{1}{\sin \vartheta} \frac{d}{d \vartheta} \left( \sin \vartheta \frac{d f_{\nu} \left( \vartheta \right)}{d \vartheta} \right) - \frac{m_{\nu}^{2}}{\sin^{2} \vartheta} f_{\nu} \left( \vartheta \right) + \lambda \left( \lambda - 1 \right) f_{\nu} \left( \vartheta \right) = 0 .
\end{equation}
So, spinor components are described by the equation well known in theory of spherical functions. It admits solutions that satisfy finiteness and unambiguity conditions, provided equality takes place
\begin{equation}
\label{det_lambda} 
\lambda \left( \lambda - 1 \right) = l \left( l + 1 \right) , 
\end{equation}
only. Here $ l $ takes positive integer values  $ l = 0,1,2,\ldots $ with $ l \geq \vert m_{\nu} \vert $, and the solutions represent associated Legendre polynomials $ P^{\vert m_{\nu} \vert}_{l} \left( \cos \vartheta \right) $. This means that eigenvalues  $ \lambda $ must satisfy equality (\ref{det_lambda}) which has two solutions 
\begin{equation}
\label{eigvL} 
\lambda_{1} = l + 1  , \quad  \lambda_{2} = - l ,
\end{equation}
one of which has positive sign, and another one - negative due to number $ l $ positiveness.

Therefore, eigenspinors  $ \hat{\Lambda} $ are spinors $ \chi_{l,M,+} $ and $ \chi_{l,M,-} $ which satisfy equation $ \hat{\Lambda} \chi_{\lambda} = \hbar \lambda \chi_{\lambda} $ with positive  $ \lambda = l + 1 $ and negative $ \lambda = - l $ eigenvalue, respectively. The spinor components, with the accuracy of constant multiplier, are expressed via Legendre polynomails 
\[
f_{1,\pm}\left( \vartheta , \varphi \right) = A_{\pm} e^{im_{1}\varphi} P_{l}^{ m_{1} } \left( \cos \vartheta \right) ,  \quad f_{2,\pm}\left( \vartheta , \varphi \right) = B_{\pm} e^{im_{2}\varphi} P_{l}^{ m_{2} } \left( \cos \vartheta \right) , 
\]
where sign $ +(-) $ corresponds to spinor with positive (negative) eigenvalue. 

Substituting these expressions into Eqs. (\ref{systL}) and requiring their identical validity, we get the  general relation for the constants:  $ A_{\pm} = \left( \lambda + m_{1} \right) B_{\pm} $  which gives 
\begin{equation}
\label{A_s,B_s} 
 A_{+} = \left(l+1+m_{1} \right)B_{+} , \quad  A_{-} = -\left(l - m_{1}\right)B_{-} , 
\end{equation}
 for each eigenstate   $ \lambda $, and the normalization condition completely determines the coefficients.  

So we conclude that eigenspinors  $ \hat{\Lambda} $ are known \cite{LandauIV} spherical spinors\footnote{Notice, normalization condition for spinors defines the incoming constants up to the phase multiplier. In Eqs. (\ref{sphsp+2}) and (\ref{sphsp-2}) it is equal to $ i^{l} $ and is introduced according to the general theory of momenta summing for $ \mathbf{\hat{J}} = \hat{\mathbf{L}} + (\hbar /2) \bm{\hat{\Sigma}} $ \cite{LandauIII}.}
\begin{equation}
\label{sphsp+2} 
\chi_{l, m_{j},+} \left( \vartheta ,\varphi \right) = i^{l} \left( \begin{array}{c} 
\sqrt{\frac{\left( l + 1 + m_{1}\right) \left( l - \mid m_{1} \mid \right)! }{4\pi \left( l + \mid m_{1} \mid \right)!}} e^{im_{1}\varphi} P_{l}^{m_{1}} \left( \cos \vartheta \right)  \\ 
\sqrt{\frac{\left( l + 1 - m_{2}\right) \left( l - \mid m_{2} \mid \right)! }{4\pi \left( l + \mid m_{2} \mid \right)!}} e^{im_{2}\varphi} P_{l}^{m_{2}} \left( \cos \vartheta \right)     
\end{array} \right) , 
\end{equation}
\begin{equation}
\label{sphsp-2} 
\chi_{l, m_{j},-} \left( \vartheta ,\varphi \right) = i^{l} \left( \begin{array}{c} 
-\sqrt{\frac{\left( l - m_{1}\right) \left( l - \mid m_{1} \mid \right)! }{4\pi \left( l + \mid m_{1} \mid \right)!}} e^{im_{1}\varphi} P_{l}^{m_{1}} \left( \cos \vartheta \right)  \\ 
\sqrt{\frac{\left( l + m_{2}\right) \left( l - \mid m_{2} \mid \right)! }{4\pi \left( l + \mid m_{2} \mid \right)!}} e^{im_{2}\varphi} P_{l}^{m_{2}} \left( \cos \vartheta \right)     
\end{array} \right)  ,
\end{equation} 
in which the numbers $m_j$ were defined in (\ref{m_1,m_2}).

\end{document}